\documentclass[11pt]{article}
\usepackage{amsmath,amssymb,amsfonts,amsthm}
\usepackage[hidelinks,breaklinks=true]{hyperref}
\usepackage{authblk}
\usepackage[english]{babel}
\usepackage[utf8]{inputenc}
\usepackage{xcolor}
\usepackage{caption}
\usepackage{sectsty}
\usepackage{subcaption}
\usepackage{graphicx}
\usepackage{dcolumn}
\usepackage{bm}
\usepackage{hyperref}

\subsectionfont{\normalfont\scshape}
\subsubsectionfont{\normalfont\itshape}

\newcommand{\R}{\mathbb{R}}
\newcommand{\C}{\mathbb{C}}

\newcommand{\Z}{\mathbb{Z}}

\newcommand{\E}{\mathcal{E}}
\newcommand{\Eb}{\bar{\mathcal{E}}}

\newcommand{\bi}{\begin{itemize}}
\newcommand{\ei}{\end{itemize}}
\newcommand{\be}{\begin{equation}}
\newcommand{\ee}{\end{equation}}
\newcommand{\re}{\text{Re}}
\newcommand{\im}{\text{Im}}

\newcommand{\beq}{\begin{eqnarray}}
\newcommand{\eeq}{\end{eqnarray}}
\newcommand{\rhomax}{\rho_{\rm MAX}}

\newcommand{\black}{\color{black}}

\newcommand\norm[1]{\left\lVert#1\right\rVert}

\usepackage{caption}
\usepackage{comment}

\def\Kasner{\kappa}

\begin{document}
\title{\bf\huge Periodic Kerr solution as an infinite soliton chain}

\author[1]{Dmitry Korotkin\footnote{dmitry.korotkin@concordia.ca}}
\author[1]{Javier Peraza\footnote{javier.perazamartiarena@concordia.ca}}
\affil[1]{Mathematics and Statistics Department, Concordia University, 1455 de Maisonneuve West, Montreal, H3G 1M8 Quebec, Canada}
\date{}

\renewcommand\Affilfont{\itshape\small}
\maketitle

\begin{abstract}
We combine numerical analysis with the inverse scattering method to study the periodic analog of   Kerr solution.  The periodic analog of the Schwarzschild' solution is  known to be regular and exhibit Kasner asymptotic behaviour for an arbitrary size of  event horizon not exceeding the period.
The previous numerical analysis of the rotating version of the periodic Schwarzschild black hole in \cite{Peraza:2022xic} based on the heat flow, together with analytical results by \cite{Peraza:2024uto} shows that there exist  obstructions to putting the  periodic Schwarzschild  solution in rotation
in a  certain parameter range.
  In this paper we apply an efficient numerical approach based on the inverse scattering method, interpreting the periodic Kerr solution as an infinite chain of solitons. This allows to completely describe the existence domain in the space  of physical parameters (the period, mass and the angular momentum);
  we study the dependence of Kasner exponent and the shape of the ergosphere on parameters of the problem.

\end{abstract}

\tableofcontents

\section{Introduction} \label{Introduction}

Over the past several decades, there have been considerably efforts in the understanding of vacuum black hole configurations with nonstandard topology of the black hole itself or the ambient space. Part of the motivation was coming from the string theory and supergravity which led to the exploration of black hole solutions in higher dimensions (e.g. \cite{Sen:1994eb, Khuri:1994gq,  Gibbons:1985ac, Larsen:1999pu, Larsen:1999pp, Myers:1987qx, Park:1995wk}). 

The first example of a vacuum black hole solution in a semi-compactified 3+1 space-time, found by Myers \cite{Myers:1987qx} (and later rediscovered in  \cite{Korotkin1994}) can be interpreted as a static black hole in a universe periodic in one spacial directions. In the other spatial directions this "periodic Schwarzschild" solution has Kasner asymptotic behaviour at infinity. The metric on the symmetry axis corresponding to this solution can be expressed in terms of $\Gamma$-functions; at all other points it is given by an explicit converging series. The geometry of this solution, was studied in \cite{Frolov:2002mq}; in particular, it was found that the horizon of the periodic Schwarzschild is not spherical but has an elongated cigar-type shape. 

In 4+1 and higher dimensions there exists a rich variety of solutions of black hole type with various topologies of the compact event horizon and the ambient spacetime (see the review \cite{Emparan:2008eg} and references therein). The higher the dimension of the spacetime is, the richer is the set of possible topologies. In 4+1, for example, besides the usual $S^3$ black ``sphere'', there are ``black rings'' $S^1 \times S^2$ \cite{Emparan:2001wn}, ``black Saturns'' \cite{Elvang:2007rd} consisting in black rings rotating around a black sphere, and various  periodic solutions \cite{Khuri:2020dbw, Khuri:2021fqu, Khuri:2022xdy} with Kasner asymptotic.

Notice that the existence of the periodic Schwarzschild solution does not contradict the Israel's theorem  \cite{Israel1968} stating the uniqueness of the static black hole, since this theorem assumes the trivial topology of the asymptotic region and the asymptotic flatness of the solution. A recent generalization of Israel's theorem was given by Reiris  \cite{Reiris2018a, Reiris2018b}, who has  shown that a metrically complete static black hole solution, without any restriction on the topology of the ambient space or the asymptotic behaviour of the metric, can be either \textit{(i)} the Schwarzschild solution, \textit{(ii)} quotients of the Rindler wedge by two independent translations, known as Boost solutions (see \cite{Reiris2018a, Reiris2018b} for details), or \textit{(iii)} finitely many black holes  on a periodic space with Kasner asymptotic. Imposing axial symmetry, it can further be proven that the third family is indeed the periodic Schwarzschild solution or its quotients  \cite{Reiris2019}.
 
The generalization of Reiris' theorem to the non-static case remains unknown. It is reasonable to expect that each static case of the Reiris' classification should have a rotating analog. The rotating analog of Schwarzschild solution is the Kerr solution. In the non-flat asymptotic sector of the classification, i.e. the Boost and the periodic Schwarzschild solutions, the rotational analog of the Kasner solution plays a central role. The rotating analogs of Kasner are solutions of the Lewis - van Stockum family \cite{Lewis, Stockum1937}, corresponding to the gravitational field of an infinitely long rotating cylinder of positive radius. It is expected that the rotating analog of the Boost solution is a special case in the Lewis - van Stockum family, since the asymptotic behaviour of the Boost can be recovered by taking the angular velocity of the cylinder to zero.

The non-trivial question which remained open until recently is whether the  "periodic Kerr" solution exists for some values of the rotation velocity, i.e.  whether  the periodic Schwarzschild solution can be put into rotation. An affirmative answer was given in \cite{Peraza:2022xic} using a numerical approach. In that work, the main idea was to use the heat flow for the harmonic map equations (the reduced Einstein Equations can be cast in a harmonic map form with domain $\mathbb{R}^3$ and target space the hyperbolic plane $\mathbb{H}^2$, \cite{Wei90}) on a half cylinder using appropriate boundary conditions in the asymptotic region, at the symmetry axis and at the horizon. 
Strong numerical evidence for the existence of the periodic Kerr solutions for a limited set of values in the space  of parameters was provided. However, the question about the complete description of the existence domain in the space of parameters (size of the black hole, period, rotation speed) remained open.

The purpose of the present article is to develop a more efficient numerical approach to the analysis of the periodic Kerr solution based on the inverse scattering method. The periodic Kerr solution is then described as an infinite soliton chain which can be analyzed numerically with higher efficiency, due to the ultra-local nature of the method, in the sense that it only depends on the point at which we are computing the solution and no other local information is require (as in a heat flow) besides the input parameters. This allows us to determine the complete existence domain in the parameter space and graph  numerically the Kasner exponent which determines the asymptotic behaviour of the solution 
(i.e., analogous to the mass parameter in the asymptotically flat case) as function of these parameters.

\begin{figure}
\centering
\includegraphics[scale=0.4]{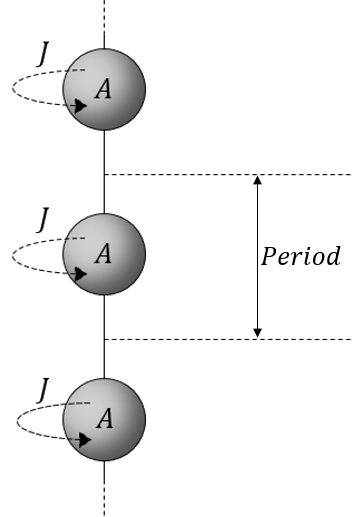}
\caption{Schematic representation of the periodic Kerr solution and the parameters.  
\label{fig_scheme_intro}}
\end{figure}

Let us describe the periodic set up and recent results in more detail. Let $2\sigma$ denote the length of event horizon in Weyl coordinates, by $L$ the period of the periodic solution along the symmetry axis, by $A$ the horizon area and by $J$ the angular momentum of the horizon. See \autoref{fig_scheme_intro} for a schematic representation.

\begin{figure}[h!]
\centering
\includegraphics[scale=0.5]{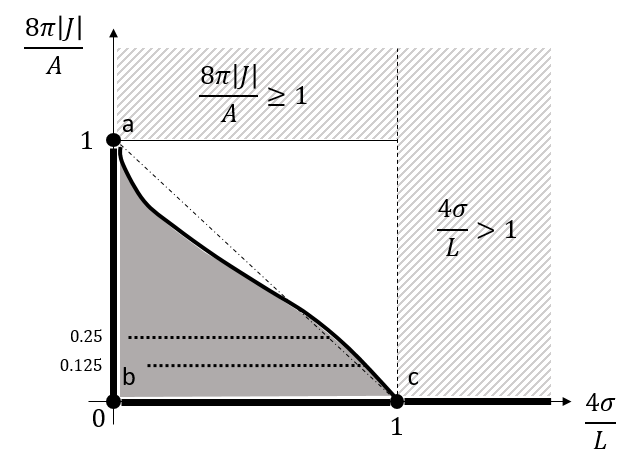}
\caption{Parameter space in terms of variables $4\sigma/L$ and $8 \pi |J| / A$. The region where non-existence of periodic Kerr was proved analytically in \cite{Dain:2011pi, Peraza:2024uto} is shown in dashed light gray. The domain where the periodic Kerr solution is shown to exist numerically is in solid grey. In the remaining domain (white) the numerical methods suggest non-existence. Points (a), (b) and (c) denote asymptotically flat extremal Kerr, Minkowski solution, and Periodic Schwarzschild with $4 \sigma = L$, respectively. In dotted line, the set of data for which existence was shown in the previous work \cite{Peraza:2022xic}.}
\label{existence_fig}
\end{figure}

In \cite{Peraza:2022xic} it was shown that the variables $\sigma/L, A$ and $J$  can be used to define the  initial data for the harmonic map heat flow corresponding to the Einstein equations. A non-trivial technical subtlety is that the asymptotic behaviour of the metric at infinity is assumed to be $z$-independent, and, therefore, must belong to the Lewis - van Stockum family. However, the corresponding Kasner exponent is unknown {\it a priori} and must be obtained as a  result of the evolution of the heat flow. In other words, the boundary conditions at the asymptotic region have to be dynamical along the heat flow. Using this method, in \cite{Peraza:2022xic} it was provided numerical evidence for the existence of periodic Kerr solutions in a discrete subset of parameters which lies within the following range:
$$
\sigma/L  < 1/4 , \quad 0 \leq J \leq A/32\pi\;.
$$
In particular, detailed runs for $A = 16 \pi$ and $J= 0.25$ and $J = 0.5$ were analyzed for various values of $\sigma/L$. The set of points
where the periodic Kerr exists is shown in dotted intervals in \autoref{existence_fig}.

In this article we analyze the full sector of the space of parameters in greater detail.
As it was shown  in \cite{Peraza:2024uto}, the condition for non-existence of solution corresponding a given point in the parameter space can be reformulated in terms of the  growth of  $f=-g_{tt}$ component of the metric
as $\rho\to \infty$. Indeed, if $f$ behaves asymptotically as $\rho^{\Kasner}$,  the Lewis' models remains complete at infinity  if $\Kasner \leq 1$; the value  $\Kasner > 1$  leads to a singularity at a finite distance from the symmetry axis.
Therefore, the  non-existence of solution with given parameters can be translated  into the  concavity test for $f$: if $f$ is concave up,  the solution is complete at infinity and vice versa. In this paper, instead of using the computationally demanding harmonic map heat flow method of \cite{Peraza:2024uto}, we compute $f$, as well as the rest of the metric,  more time-efficiently by interpreting periodic Kerr as an infinite soliton chain. 

The theory of solitons was applied to the stationary axially symmetric Einstein equation in   70's. The existence of  an infinite-dimensional group of symmetries was first discovered by Geroch  \cite{Geroch:1970nt, Geroch:1972yt}. 
The zero curvature representation (the "Lax representation") for these equations was found in 1978 by Belinskii and Zakharov \cite{Belinsky:1978nt, Belinsky:1979mh} and Maison \cite{Maison:1978es};
these representations, although equivalent, have a rather different form; another (again equivalent) zero curvature  representation was found later in  \cite{Neugebauer:1980}. Here we use the  latter 
zero curvature  representation.

Using the zero curvature  representation one can  add an arbitrary number of solitons to a given "seed" solution.
In this paper we use this method iteratively, by adding two solitons with suitable parameters at a time with the aim of describing an infinite soliton chain in the limit.
Then  we  get a sequence of solutions, $\E_{n}$, indexed by the number of (pairs of ) solitons, and study numerically the convergence properties in the limit $n \rightarrow +\infty$. In particular, we find that the convergence is almost logarithmic, with velocity close to 1 (see \autoref{sec_results}). This procedure is   a natural non-linear generalization   of the linear superposition principle used to derive the  periodic Schwarzschild solution \cite{Myers:1987qx,Korotkin:1994cp}.

The periodic Schwarzschild solution exists for any values of $\sigma$ and $L$ as long as $2\sigma<L$. However, not every periodic Schwarzschild solution can be put in rotation and for those which can there exists an upper bound 
on the angular momentum which depends non-trivially on $\sigma$ and $L$.
It is a non-trivial problem to  fully describe the parameter space of the possible periodic Kerr configurations. 
There exist two obvious restrictions on the range of  parameters for such. First, to avoid horizon overlap, we assume $2\sigma < L$.
Second, as any compact horizon satisfies the inequality $|J| \leq A/8\pi$ \cite{Dain:2011pi}, this condition gives an upper bound for the maximum angular momentum in terms of the area. 
Moreover, in the recent paper \cite{Peraza:2024uto} there  was shown that the periodic analogue of Kerr solution does not exist if  $4\sigma  > L$ i.e. if the length of event horizon in Weyl coordinates exceeds half of the period.    This result was proven in  \cite{Peraza:2024uto} by pure analytic methods, without assuming any asymptotic behaviour other than the topology of the asymptotic end (i.e., a cylinder $\R \times \mathbb{T}^2$). 

Thus, these analytic results restrict the possible existence domain  to the rectangle
$4\sigma/L \leq 1$, $8 \pi |J| / A \leq 1$. In this paper we further restrict  the  domain of existence. 

It is convenient to parametrize the family of asymptotically flat Kerr solutions by $\sigma$ and an auxiliary real parameter $0\leq p \leq 1$. We also denote $q=\sqrt{1- p^2}$ and $\alpha=p + iq$. The set  of parameters $(\sigma,L,p)$ is more suitable for the soliton theory approach and is used in this paper. In terms of these parameters one can express the mass $M$, area $A$ and the angular momentum $J$ as follows:
\be \label{variables_sigma_p_q}
M=\frac{\sigma}{p}\;,\hskip0.7cm 
J=\frac{q}{p^2}\sigma^2\;,\hskip0.7cm A= 8\pi M(M+\sigma)\;.
\ee

In the  limit $p \rightarrow 1$ we have $J \rightarrow 0$   and in the limit $p \rightarrow 0$ we have $|J| \rightarrow A/8\pi$. 


The first main result of this paper is the complete numerical description of the existence domain of periodic Kerr solution. 
%
%
The existence domain is shown in \autoref{existence_fig} in terms of parameters  $4\sigma/L$ and $8\pi |J|/A$
 ( due to scaling invariance this is sufficient to describe the domains in the  three-dimensional space of parameters $(\sigma,L,J)$).




As a reference, in \autoref{existence_fig} we indicate some special solutions (the points (a), (b), (c) and thicker lines). The horizontal bold black segment indicates the periodic Schwarzschild solutions. The vertical bold black segment corresponds to asymptotically flat solutions  $L= +\infty$ (recall that $\sigma > 0$ if $J<2$). Thus, points (a) and (b) correspond to the asymptotically flat extreme Kerr and Schwarzschild solutions. Point (c) is the periodic Schwarzschild solutions at the boundary between those that can be put into rotation and those which can not. 

\begin{figure}[h!]
\centering
\includegraphics[scale=0.35]{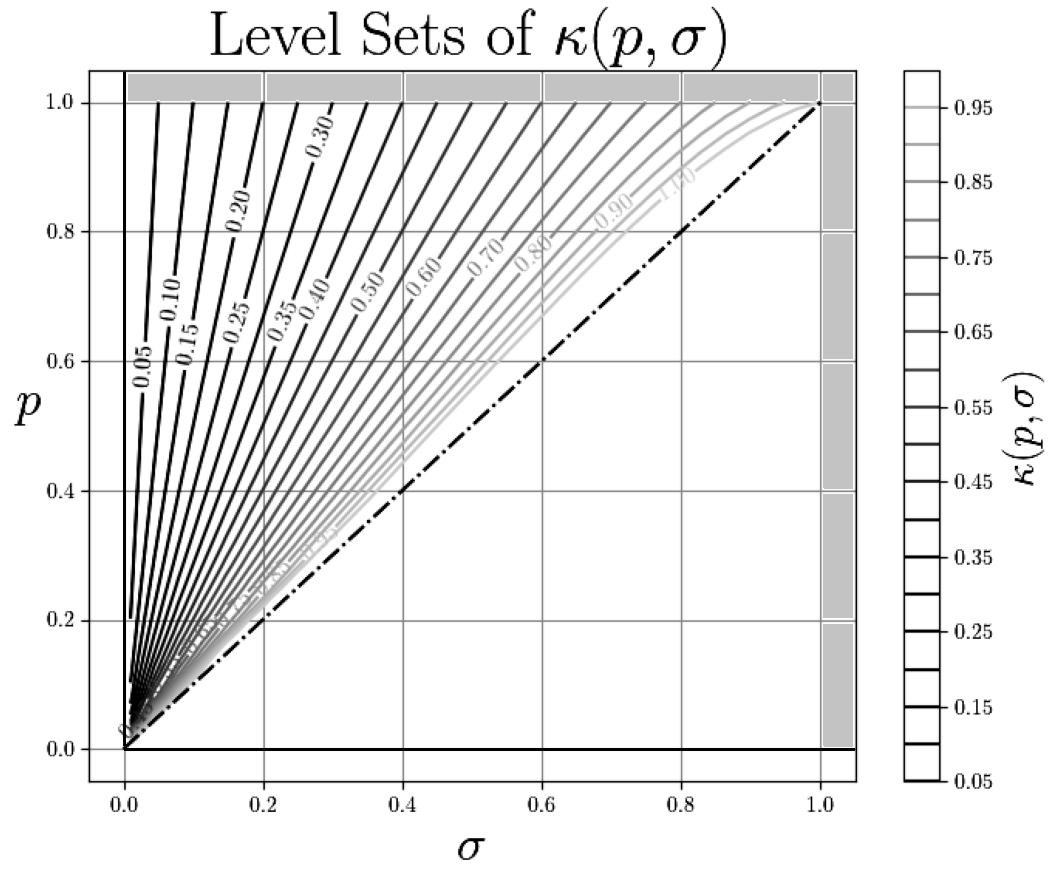}
\caption{Kasner exponent $\Kasner$ in the domain of existence of the periodic Kerr solution. We fixed $L=4$ for a better diagram.
\label{Figure_level_sets_3D}. }
\end{figure}

The second main result is the graph in \autoref{Figure_level_sets_3D}, corresponding to the Kasner exponent $\Kasner$ of the periodic Kerr solutions in the existence domain. For $p=1$ (periodic Schwarzschild) this graph is in agreement with the analytic formula $\Kasner=4\sigma/L$. Small deviations from the heuristic formula $\Kasner = \frac{4 \sigma}{p L}$ (based on the periodic Schwarzschild case)  can be noticed as we get closer to the line $p = \sigma$.  This was already shown in \cite{Peraza:2022xic} when comparing the value of $M$ between the periodic Kerr and periodic Schwarzschild with the same parameter $\sigma$. Observe that the domain of existence is such that we have inequality
\be 
\kappa \geq \frac{4\sigma}{pL}.
\ee 
This inequality can be understood in terms of the rotational energy available to the black hole. As a consistency check, we verify that all constructed solutions verify the Penrose-like inequality,
\be
M \geq \sqrt{\frac{A}{16 \pi } + \frac{4 \pi J^2}{A}} ,
\ee 
with equality approached in the asymptotically flat limit $p \rightarrow 1$ and $\sigma \rightarrow 0$.

%
The article is organized as follows. In \autoref{sec_preliminaries}, we remind the construction of periodic Schwarzschild solution and review the inverse scattering method for Ernst equation. We describe the iterative approach which is carried out by adding two solitons to a given  background  and taking the number of solitons to infinity. In \autoref{sec_setup}, we present the numerical scheme and discuss the convergence of the iterative procedure. In \autoref{sec_results} we compute   Ernst potential and metric coefficients of periodic Kerr solution numerically. Here we also compare our current approach to the heat flow evolution method of \cite{Peraza:2022xic} and compute  the Kasner exponent numerically in some test cases. In \autoref{sec_existence} we
analyze  the domain of existence of the periodic Kerr solution. We also  graph the Kasner exponent in the existence domain.  We compute the critical value of the ratio of the size of the static  black hole
to the period at which no rotation is possible, and study how the shape  of the ergosphere depends on the angular  momentum for a given period on the size of the event horizon. Finally, in \autoref{sec_outlook} we discuss possible directions of further research.


\section{Integrability of stationary axisymmetric gravity} \label{sec_preliminaries}

\subsection{Metric in Weyl coordinates and Ernst equation}

First, remind that 
the stationary axially symmetric Einstein equations can be formulated using  the canonical Weyl coordinates (cf. \cite{Wald1984}) denoted by $(t,\phi , \rho, z)$, such that $\partial_t$ and $\partial_\phi$ are the time-like and rotational Killing fields, respectively, and $\rho$ and $z$ are the remaining coordinates; the metric coefficients depend only on $\rho$ and $z$. In these coordinates the line element takes the form
\begin{equation} \label{ernst_metric_form}
ds^2 = f^{-1} (e^{2k}(dz^2 + d\rho^2) + \rho^2 d \phi ) - f (dt + F d\phi)^2
\end{equation} 
where $f,k$ and $F$ are functions depending only on $(\rho, z)$. The periodicity of the solution along the $z$ coordinate with period $L>0$ means the invariance of the metric under the translation $z\to z + L$. Since we have two periodic directions, one along $\partial_z$ and the other along $\partial_\phi$ due to the axisymmetry, the asymptotic end of the corresponding manifold is a topological cylinder $\mathbb{R}_+ \times \mathbb{T}^2$. The coordinates associated to the $T^2$ factor are $\phi$ and $z$, while $\mathbb{R}_+$ is parametrized by the coordinate $\rho$\footnote{The unboundedness of $\rho$ is a non-trivial feature of Weyl coordinates, since $\rho$ is a harmonic function on the original three dimensional manifold. We refer the reader to a thorough analysis of this issue in \cite{Reiris2018b} for the static case, extendable to the stationary case.}.

For black hole solutions  the Weyl coordinates degenerate at the horizon which then looks like  a segment of the symmetry axis $\rho =0$.
However, the norm of $\partial_\phi$ remains non-vanishing on the horizon. Let us denote the length of the horizon in the coordinate $z$ by $2\sigma$; then  the area of the horizon is given by
\begin{equation}
A := 2 \pi \int_{-\sigma}^\sigma | e^{k} F | dz.
\end{equation}

Weyl coordinates are well-suited for numerical analysis, since the degeneracy of the metric at the horizon can be controlled analytically \cite{Wei90, Dain2009}. The three natural  variables defining a periodic Kerr solution can be chosen as   the ratio $\sigma/L$, the area $A$ of the horizon and the  angular momentum, $J$, which is  defined as the Komar integral associated to the rotational Killing field $\partial_\phi$:
\begin{equation} \label{Komar_ang_mom}
J := \frac{1}{16 \pi} \int_\Sigma \epsilon_{\mu \nu \alpha \beta } \nabla^\alpha (\partial_\phi)^\beta \ . 
\end{equation} 
Here $\Sigma$ some closed surface surrounding the horizon\footnote{The closed surface can be a sphere or a torus in the periodic setup}, $\nabla$ is the Levi-Civita connection of the induced metric on $\Sigma$ and $\epsilon$ is the volume element.

Given the line element as in \eqref{ernst_metric_form}, Einstein equations reduce to two mixed elliptical equations for $f$ and $F$, plus  quadrature equations for $k$ and $F$. An equivalent prescription to solve Einstein equations is given by the well-known Ernst equation \cite{Ernst:1967wx},
\be  \label{Ernst_equation}
(\E + \Eb) ( \E_{zz} + \frac{1}{\rho} \E_\rho +  \E_{\rho\rho}) = 2 ( \E_z^2 +  \E_\rho^2)\;,
\ee
where $\E$ (Ernst potential) is a complex valued function depending on $(\rho,z)$. It contains the information regarding the elliptical system for $f$ and $F$. The components of the metric can be recovered by the identity
\be 
\E = f + i \varpi\;,
\ee
where $\varpi$ is the twist potential for the stationary Killing field $\partial_t$. The quadratures for $k$ and  $F$ are expressed in terms of $\E$ as follows
\be \label{conformal_factor}
  F_\xi = i \rho\frac{\partial_\xi  \im \E}{f^2}  , \quad   k_\xi  =  i\rho\frac{\E_\xi  \bar{\E}_\xi}{2 f^2}  \;,
\ee
where we denote $\xi := z+ i\rho$.

If $\E$ is real, then \eqref{Ernst_equation} can be rewritten as a linear equation
\be 
\label{linear_eq}
(\ln\E)_{zz} + \frac{1}{\rho} (\ln\E)_\rho +  (\ln\E)_{\rho\rho}= 0\;,
\ee 
This case corresponds to static  axisymmetric solutions, when $F= 0$ (cf. \eqref{conformal_factor}). 

\subsection{Static (linear) case: periodic Schwarzschild}

The asymptotically flat Schwarzschild solution, depends on  only one natural parameter, the mass $M$. This parameter is related to the area of the event horizon by the simple formula, $A = 16 \pi M^2$. The mass gives an unambiguous solution given the normalization at the asymptotic region of the $g_{tt}$ component of the metric to $-1$.

In Weyl-Papapetrou coordinates, as we mentioned before, the horizon becomes a segment on the symmetry axis $\rho = 0$ of  length $2\sigma \equiv 2 M$. 

To define the periodic Schwarzschild solution in Weyl coordinates, one introduces an extra parameter, the period $L$ in $z$-coordinate. This quantity does not have a direct physical meaning, since it is a \textit{coordinate} parameter. 

Given a pair $(\sigma , L)$, we can construct  the periodic analogue of Schwarzschild solution, by means of an exact analytic series \cite{Korotkin1994}. Here we briefly review the construction.

The Ernst potential  $ \E$ 
for the Schwarzschild solution is real and given by
$$
\E_S = \left( \frac{\sqrt{(z-M)^2 + \rho^2} +  \sqrt{(z+M)^2 + \rho^2} - 2M}{\sqrt{(z-M)^2 + \rho^2} +  \sqrt{(z+M)^2 + \rho^2} + 2M} \right)^2.
$$
Denote $U_S:=\frac{1}{2} \ln \E_S$.
Since the Ernst equation  is linear (\ref{linear_eq})  in the static case, 
we can take a linear superposition of  an arbitrary finite number of solutions to obtain a new solution. 
This property, together with the translation invariance of solutions (if $U(\rho,z)$ is a solution, then so is $U(\rho, z+ L)$), is used for the construction of the periodic solution. Namely, let $L$ be 
such that two consecutive horizons do not overlap, i.e. $L > 2 M$. Then, the function $U$ for the periodic Schwarzschild solution is constructed as follows:
\be \label{U_PS}
U_{PS} = U_S(\rho ,z) + \sum_{n=1}^\infty \left( U_S(\rho ,z + nL) + U_S(\rho ,z - nL) + \frac{4M}{nL} \right),
\ee
where the counter-terms $\frac{4M}{nL}$ is introduced to regularize the series. This solution (\ref{U_PS}) turns out to be  regular everywhere outside of the event horizon.
The leading term of the asymptotic behaviour of the solution (\ref{U_PS}) as  $\rho\to\infty$ is given by the  Kasner solution,
\be 
ds^2_K \approx - c_0 \rho^{\Kasner} dt^2 +  c_1 \rho^{\Kasner^2/2 - \Kasner} (dz^2 + d\rho^2) + c_2 \rho^{2 - \Kasner} d\phi^2\;
\label{Kasner}
\ee
for some constants $c_0,c_1$ and $c_2$, and with the Kasner exponent given by 
\be 
\Kasner  = 4\frac{\sigma}{L}\;.
\label{KsL}
\ee

The Kasner exponent $\Kasner$ can be related to the mass parameter $M$, via a fixing of the residual gauge, consisting in scaling of $\partial_t$ vector. Indeed, the fixing of the $g_{tt}$ component of the metric is related to the surface gravity $\tau$ of the horizon (for a complete discussion on the definition of surface gravity we refer the reader to \cite{Jacobson:1993pf}). Smarr's formula in the static and asymptotically flat case gives
\be \label{Smarr_static}
M = \frac{\tau A}{4\pi}\;.
\ee
Since Smarr's identity still holds if the Komar integrals evaluated at the torus at constant $(\rho,t)$ slices, we have 
$$
M(\sigma ,L) = \frac{\tau(\sigma ,L) A(\sigma ,L)}{4\pi}\;.
$$
 It can be shown that the relation $\tau(\sigma ,L) A(\sigma ,L) = 4\pi \sigma$ holds in Weyl coordinates, and therefore we have $M(\sigma ,L) = \sigma$. On the other hand, the Kasner asymptotic behaviour of the solution means that the fixing of $g_{tt}$ to tend  asymptotically $-1$ is no longer possible. Instead, we can use the freedom in the scaling of $\partial_t$ to fix the surface gravity $\tau$ of the horizon to be equal to the surface gravity  of the asymptotically flat Schwarzschild with parameter $M$.

Numerical analysis of this solution was performed in \cite{Frolov2003}, where the authors focused on the distortions of the horizon of the black hole due to the periodic topology. To construct a solution, the authors presented two approaches. They use the series representation of the solution, given in \eqref{U_PS} as well as an integral representation via the Green function of an infinite set of ``rods'' (i.e., segments of length $2\sigma$) positioned periodically along the $z$-axis. 

The authors of \cite{Frolov2003} calculate the geometrical features of the  horizon (size, shape and isometric embeddings into three dimensional Euclidean space) and its thermodynamic quantities such as the proper length of the horizon, redshift factor, and surface gravity, showing that the black hole horizon becomes elongated in the direction of the symmetry axis  due to the periodic topology. 
Stability of periodic Schwarzschild is  discussed also discussed in \cite{Frolov2003} from the energy point of view. The results in \cite{Frolov2003} provide a semi-analytic framework to understand the non-trivial effects of compactification on black hole geometry and thermodynamics. 

In this paper, we will also focus on the distortions of the horizon geometry due to the periodic topology in the periodic Kerr case, and graph the three dimensional  isometric embeddings of both the horizon and the ergospheres (see \autoref{sub_sec_ergosphere}).

\subsection{Stationary case: integrability and auxiliary linear system}

The  Ernst equation \eqref{Ernst_equation} is the compatibility condition for the following linear system \cite{Neugebauer:1979iw, Neugebauer:1980},
\beq \label{lax_1}
\frac{\partial\Psi}{\partial \xi} &=&\left[ \left( \begin{array}{cc}
A & 0 \\
0 & B
\end{array} \right) + \sqrt{\frac{\lambda - \bar{\xi}}{\lambda - \xi}} \left( \begin{array}{cc}
0 & A \\
B & 0
\end{array} \right) \right]\Psi,   \\
\frac{\partial\Psi }{\partial \bar{\xi}}&=& \left[\left( \begin{array}{cc}
\bar{B} & 0 \\
0 & \bar{A}
\end{array} \right) + \sqrt{\frac{\lambda - \xi}{\lambda - \bar{\xi}}} \left( \begin{array}{cc}
0 & \bar{B} \\
\bar{A} & 0
\end{array} \right)\right] \Psi, \label{lax_2}
\eeq
where $\lambda \in \C$ is known as the \textit{constant} spectral parameter \cite{Maison:1978es}, $\Psi(\lambda , \xi)$ is a $2\times 2$ matrix and 
\be 
A = \frac{\partial_\xi \E}{\E + \bar{\E}} , \quad B = \frac{\partial_\xi \bar{\E}}{\E + \bar{\E}}.
\ee
The constant spectral parameter $\lambda$ is related to the \textit{variable} spectral parameter $\gamma$ by equation
\be \label{two-sheets_surface}
\gamma = \frac{2}{\xi - \bar{\xi}} \left(\lambda - \frac{\xi + \bar{\xi}}{2} + \sqrt{(\lambda - \xi)(\lambda - \bar{\xi})} \right)\;.
\ee

To recover the physical metric from $\Psi$, we have to suitable normalize the function $\Psi$. We will work with the following normalization at $\lambda = +\infty$ (corresponding to $\gamma = \infty$):
\be
\Psi(\lambda = +\infty )\left(\begin{matrix} 1 \\ -1\end{matrix} \right)= \left(\begin{matrix} 1 \\ -1\end{matrix} \right)
\label{normcon}\ee
which implies
\be \label{normalization_Psi}
\Psi(\lambda = +\infty ) = \frac{1}{2} \left( \begin{matrix}
1 + \E & - 1 + \E\\
- 1 + \Eb & 1 + \Eb
\end{matrix} \right)\;.
\ee  
That   implies the following formula for the  Ernst potential
\be \label{potential_eq}
\frac{\E + 1}{\E - 1} = \frac{\Psi_{11}}{\Psi_{12}}\;.
\ee

The matrix $\Psi$ is assumed to satisfy the following reality condition, 
$$
\overline{\Psi (\bar{\lambda})} = \sigma_1 \Psi(\lambda) \sigma_1,
$$
and the involution condition, 
$$
\Psi (\lambda^*) = \sigma_3 \Psi(\lambda) \sigma_3,
$$
with $*$ interchanging the sheets of the Riemann surface ${\mathcal L}$ of the function  $\sqrt{(\lambda - \bar{\xi})(\lambda - \xi}$. 

Addition of a combination of solitons to a given background solution can be  described  as a multiplication of $\Psi$ from the left by a function meromorphic on the Riemann surface 
${\mathcal L}$
 \cite{Belinsky:1978nt, Neugebauer:1980}. 
The transformation of metric coefficients $f$ and $F$ under  addition of  a finite number of solitons to a background solution was given in \cite{Belinsky:1978nt,Kramer1986}.

The real part of the Ernst potential is related to $\Psi$ via the normalization prescription \eqref{normalization_Psi},
\be \label{det_eq}
\re \E = \det \Psi (\lambda = + \infty)\;,
\ee
and, therefore, the coefficient $f ( = \re \E)$ form the physical metric can be computed directly from $\Psi$.

With the normalization \eqref{normalization_Psi}, we can compute the metric coefficient $F$ by means of the identity
$$
\partial_\xi (\Psi^{-1} \partial_{1/\lambda} \Psi ) = \Psi^{-1} \partial_{1/\lambda} \left( \partial_\xi \Psi \Psi^{-1} \right) \Psi\;.
$$

In  the limit $\lambda \rightarrow +\infty$ we get
$$
\partial_\xi (\Psi^{-1} \partial_{1/\lambda} \Psi ) \mid_{\lambda = +\infty} =  \frac{ i \rho}{2(\E + \Eb) } \left( \begin{matrix}
(\Eb^2 -1) A + (1 - \E^2) B & (1 + \Eb)^2 A - ( \E - 1)^2 B \\
( \E + 1)^2 B  - ( \Eb - 1)^2 A &  (1 - \Eb^2 ) A + (\E^2 - 1) B
\end{matrix} \right).
$$
Using the trace operator $\text{tr} \left(\cdot V \right)$, with 
$V = \left( \begin{matrix}
2 & 1 \\
-1 & 0
\end{matrix} \right)$, 
we can integrate the equation for $F$ in terms of $\Psi^{-1} \partial_{1/\lambda} \Psi$ to get,
\be \label{component_F_metric}
F= c + \frac{i}{4} \text{Tr} \left( \Psi^{-1} \partial_{1/\lambda} \Psi V \right) \big|_{\lambda = +\infty},
\ee
for some constant $c$. This constant is defined by imposing the condition $F(0,z)=0$ on the symmetry axis outside horizons, see e.g. \cite{Neugebauer2009}.

In this work, we are interested in the limit where an infinite number of identical, equidistant, and coaxial horizons are successively added, thus resulting in a configuration corresponding to a chain of infinitely many stationary black holes. This construction leads to a ``periodic Kerr'' solution, as described in \cite{Korotkin:1994cp}. 

This iterative approach involves the successive introduction of two solitons to a background spacetime solution as the primary building block for generating the periodic structure. In the next subsection we review the procedure by which a new pair of solitons is added to a background solution.

\subsection{Adding two solitons to a background solution}

In this subsection we review the solution generating method when two solitons are superposed with  a background solution. This will be  the basic step for our iterative numerical computations.

Denote by $w(\lambda , \xi) = \sqrt{(\lambda - \xi)(\lambda - \bar{\xi})}$, and assume that $\Psi_0$ is  a background solution to the linear system \eqref{lax_1},\eqref{lax_2} which corresponds to the Ernst potential $\E_0$. We are adding two new solitons, at positions $\lambda_1 , \lambda_2 \in \R$ and with associated unitary constants $\alpha_1$ and $\alpha_2$ respectively,  $| \alpha_i | = 1$, for $i=1,2$. In the particular case of adding solitons that correspond to a stationary black holes, the constants $\alpha_1$ and $\alpha_2$ should be  related by the formula
$$
\alpha_1 \alpha_2 = -1\;.
$$

The normalization condition (\ref{normcon})  implies that the Ernst potential for the background solution can be obtained from the value of $\Psi_0$ at $\lambda=\infty$
$$
\Psi_0(\lambda = +\infty) = \frac{1}{2} \left( \begin{matrix}
1 + \E_0 & - 1 + \E_0 \\
- 1 + \Eb_0 & 1 + \Eb_0
\end{matrix} \right).
$$ 

Let  $\Psi$ be the new solution to the linear system \eqref{lax_1},\eqref{lax_2} defined by 
\be \label{def_Psi_new}
\Psi(\lambda,\xi) = T(\lambda,\xi) \Psi_0(\lambda, \xi),
\ee
where $T$ is a two-by-two matrix whose components are polynomials in $\lambda$ and $w$ with coefficients being functions of $\xi$. The reality and involution conditions imply the following form of  the matrix $T$ corresponding to two solitons on the given background:
$$
T(\lambda,\xi) = \frac{1}{\lambda} \left( \begin{array}{cc}
Q_1(\lambda) & w(\lambda , \xi) P_0 (\lambda) \\
w(\lambda , \xi) \overline{P_0 (\bar{\lambda})} & \overline{Q_1 (\bar{\lambda})}
\end{array} \right),
$$
where $Q_1$ and $P_0$ are complex polynomials of degree 1 and 0, respectively.
The  matrix $\Psi$ is assumed to satisfy the normalization condition (\ref{normcon}) and its zero eigenvectors at $\lambda_j$ are assumed to have the form
\be \label{eigen}
\Psi (\lambda_i) \left( \begin{matrix}
1 \\
\alpha_i 
\end{matrix} \right) = \left(\begin{matrix}
0 \\
0
\end{matrix} \right).
\ee
for constant $\alpha_1$ and $
\alpha_2$.
The condition (\ref{normcon})  implies a relation between the highest degree coefficients of polynomials $Q_1$ and $P_0$,
 $$
q_1 - p_0 = 1\;.
$$
The equations (\ref{eigen})  can be written as 
\be  \label{betas_eq}
T (\lambda_i, \xi) \left( \begin{array}{c}
1 \\
\beta_i
\end{array} \right) = 0, 
\ee 
with $\beta_i$ defined in terms of  constants $\alpha_i$ and the background solution $\Psi_0$,
\be \label{leading_coef_condition}
\beta_i := \frac{\mu_i}{\nu_i}, \hskip1.0cm \left( \begin{array}{c}
\nu_i \\
\mu_i
\end{array} \right) = \Psi_0 (\lambda_i, \xi) \left( \begin{array}{c}
1 \\
\alpha_i
\end{array} \right).
\ee

Denote the coefficients of $Q_1(\lambda)$ and $P_0(\lambda)$ as follows 
$$
Q_1(\lambda) = q_1 \lambda + q_0 , \quad P_0(\lambda) = p_0\;.
$$
Then equation \eqref{betas_eq} and condition \eqref{leading_coef_condition} give
\beq \label{q_p_ceof_1}
0 &=& q_0 + q_1 \lambda_1 + w(\lambda_1) \beta_1 p_0\;, \\
0 &=& q_0 + q_1 \lambda_2 + w(\lambda_2) \beta_2 p_0\;, \label{q_p_ceof_2}\\
1 &=& q_1 -p_0\;, 
\label{q_p_ceof_3}
\eeq
where $\beta_i$ are given by \eqref{betas_eq}. The solution to  equations \eqref{q_p_ceof_1}, \eqref{q_p_ceof_2} and \eqref{q_p_ceof_3} is given by
$$
q_1 =  \frac{w(\lambda_1) \beta_1 - w(\lambda_2) \beta_2}{\lambda_1 - \lambda_2 + w(\lambda_1) \beta_1 - w(\lambda_2) \beta_2} \;,
$$
$$
p_0 = - \frac{\lambda_1 - \lambda_2}{\lambda_1 - \lambda_2 + w(\lambda_1) \beta_1 - w(\lambda_2) \beta_2} \;,
$$
$$
q_0 =\frac{w(\lambda_2) \beta_2 \lambda_1 - w(\lambda_1) \beta_1 \lambda_2}{\lambda_1 - \lambda_2 + w(\lambda_1) \beta_1 - w(\lambda_2) \beta_2} \;. 
$$
The corresponding  Ernst potential found computed using \eqref{def_Psi_new} and \eqref{potential_eq} is given by
\be 
\label{Ernst_plus_two_solitons}
\E = q_1 \E_0 + p_0 \Eb_0\;.
\ee

\subsubsection{Exact formulas for the metric coefficients} \label{exact_formulas_coeff}

Let $f_0,F_0$ and $k_0$ be the  metric coefficients corresponding to the background solution, cf. \eqref{ernst_metric_form}. The function $f=\re \E$ can be found by equation \eqref{det_eq}:
\be \label{f_from_f_0}
f = (q_1 \bar{q}_1 - p_0 \bar{p}_0) f_0\;.
\ee 

The function $F$ is related to  $F_0$ via equation \eqref{component_F_metric},

\beq \label{F_from_F_0}
F &=& c_{new} + F_0 + \frac{i}{4} \text{Tr} \left( T^{-1} \partial_{1/\lambda} T \Psi_0 \left( \begin{matrix}
2 & 1 \\
-1 & 0
\end{matrix} \right) \Psi_0^{-1} \right) \bigg|_{\lambda = +\infty}
\eeq
where we use the cyclic properties of the trace, and the constant $c_{F,new}$ is fixed by condition of regularity of the new metric at the axis of symmetry. 
For Kerr solution $F$ is given by \autoref{Kerr_explicit}. 

The formula for the new conformal factor $k$ in terms of the background conformal factor $k_0$ looks as follows \cite{Kramer1986}

\be \label{k_from_k_0}
e^{2k} = c_{k,new} \frac{e^{2k_0}}{4 f_0^2} \frac{1}{s_1 s_2} \det \left( \begin{matrix}
\nu_1 & s_1 \mu_1 \\
\nu_2 & s_2 \mu_2
\end{matrix} \right) \det \left( \begin{matrix}
\mu_1 & s_1 \nu_1 \\
\mu_2 & s_2 \nu_2
\end{matrix} \right),
\ee
where $\mu_i$ and $\nu_i$ are given in \eqref{betas_eq} and $s_i := \sqrt{\frac{\lambda_i - \xi}{\lambda_i - \bar{\xi}}}$. The constant $c_{k,new}$ is  fixed so that no conical singularities arise at the axis of symmetry.

Overall, there is a scaling constant that is not fixed by the procedure described so far, which is the residual gauge from scalings of $\partial_t$. This is related to the stationary Killing field scaling,
\be 
t \mapsto \tilde{t} = \gamma t, \quad \partial_t \mapsto \frac{1}{\gamma} \partial_{\tilde{t}},
\ee 
which in turn implies a scaling of the Ernst potential
\be \label{scaling_E}
\E \mapsto C_{\gamma} \E,
\ee
for some constant $C_{\gamma}$ that depends on $\gamma$. The underlying residual gauge freedom is an artifact of the particular coordinate choice \cite{Peraza:2022xic, Peraza:2023boa}. Usually, this residual gauge freedom is lifted by a natural boundary condition at the asymptotic region, that is, imposing that the modulus of the stationary Killing vector converges to $-1$ as $\rho \rightarrow + \infty$. In the periodic case, the solutions are asymptotically Kasner \cite{Peraza:2023boa} and therefore one can no longer fix the modulus to a particular value. We will discuss different prescriptions for fixing this arbitrary constant in \autoref{sub_sec_iterative_cons}.

\subsubsection{Ergospheres and singularities} \label{erg_sub}

The ergosphere is defined by the condition $g_{tt}=0$ which is equivalent to vanishing as  of the real part of the Ernst potential,
\be
E_f := \{ f = 0, \quad \rho > 0 \}.
\label{ergo}
\ee
The   ergosphere of  Kerr solution surrounds horizons; it  is defined by equation (\ref{ergo}).

Although equations for metric coefficients are singular on $E_f$, cf. \eqref{conformal_factor}, it can be shown (see e.g. \cite{Chrusciel:2006se}) that the corresponding metric  is smooth on  $E_f$.

When constructing a new solution by adding two solitons to a background solution \textit{without} the ergosphere is given by  \eqref{f_from_f_0} i.e.
\be
E_f = \{ q_1 \bar{q}_1 - p_0 \bar{p}_0 = 0, \quad \rho > 0 \}
\label{ergeq}
\ee
since $f_0 \neq 0$ when $\rho > 0$. However, when adding two solitons to a background solutions that \textit{already} has an ergosphere (given by set $E_{f_0}$), the right-hand side  in \eqref{f_from_f_0} vanishes on $E_{f_0}$ unless the function $q_1 \bar{q}_1 - p_0 \bar{p}_0$ is also singular there. As can be seen   from the four-soliton solution, and inductively from the equations defining $q_1$ and $p_0$, the term $q_1 \bar{q}_1 - p_0 \bar{p}_0$ is singular in the set $E_{f_0}$, such that the product in the right hand side of \eqref{f_from_f_0} is regular and non-vanishing on $E_{f_0}$.  Therefore, the ergosphere of the new solution is always defined given by the equation (\ref{ergeq}).

Numerically, equation \eqref{f_from_f_0} poses a difficulty for iterative solution generating techniques, which translates into the well-known floating-point arithmetic problems related to  division of two numbers close to zero. The precise structure of singularities and zeros for $q_1 \bar{q}_1 - p_0 \bar{p}_0$ can be treated by avoiding numerical computations where analytic treatment can be performed.

\section{Setup of the Periodic Problem} \label{sec_setup}

In this section we define the numerical problem, which is used as an iterative approximation to a periodic solution using the inverse scattering method. The numerical nature of our analysis implies that 
the approximate solution is constructed   on a finite domain.

\subsection{Iterative construction of solutions} \label{sub_sec_iterative_cons}

\begin{figure}[h!]
\centering
\includegraphics[scale=0.5]{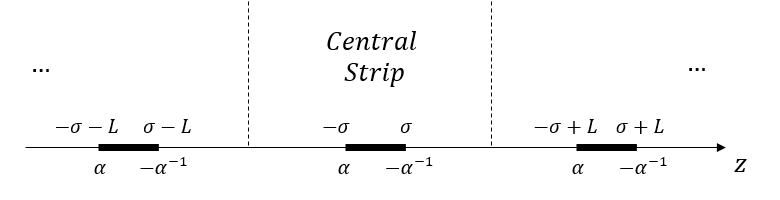}
\caption{\label{solitons_diagram} Location of solitons and their respective constants in $(\rho,z)$ coordinates. The central strip located between the dashed lines is the fundamental domain of the periodic solution. }
\end{figure}

Our aim is to construct periodic analogues of Kerr solution, which can be interpreted  as an infinite superposition of identical and equidistant Kerr black holes along the $z$-axis. Periodicity  of the metric along $z$-axis with period  $L$  implies the following periodicity for function $\Psi$: 
$$
\Psi (\lambda + L , \xi + L) = \Psi(\lambda , \xi) R(\lambda)
$$
where the  matrix $R(\lambda)$ is independent of $(\rho ,z)$. Then, to construct $\Psi$, we consider  an infinite-soliton solution with  zeros of $\det \Psi$  located at the points 
$$
\{\lambda_1 + mL\}_{m\in \Z}\cup \{\lambda_2 + m L\}_{m\in \Z}\;,
$$
such that the  zero eigenvectors of $\Psi$ at   $\lambda_1 + mL$ are given by $(1,\alpha)^t$ for all $m$ and the  zero eigenvectors of $\Psi$ at   $\lambda_2 + mL$ are given by $(1,-\alpha^{-1})^t$ for all $m$.
 In the sequel we choose  $\lambda_1 = -\lambda_2 = - \sigma$, see \autoref{solitons_diagram}. Therefore, the three parameters used as an input are $(\sigma , L , \alpha)$.

A constructive way to obtain a periodic solution is by adding solitons in an iterative procedure. For each $n \geq 1$, we consider the solution $\Psi_n$ to the linear system \eqref{lax_1},\eqref{lax_2} with zeros of $\det \Psi_n$ located at
\be 
\{-\sigma + mL\}_{m=-n}^{n}\cup \{\sigma + m L\}_{m=-n}^{n},
\ee
with parameters $\alpha$ and $-\alpha^{-1}$ defining the eigenvectors at $-\sigma + mL$ and $\{\sigma + m L\}$, respectively.
The corresponding Ernst potential computed from $\Psi_n$ is denoted by $\E_n$, and is a $4n+2$-soliton solution for the given parameters. 

Using \eqref{Ernst_plus_two_solitons}, we get the following recursive relation between $\E_n$ and $\E_{n+1}$,
\be \label{E_n+1_after_E_n}
\E_{n+1} = q_1^{+} (q_1^{-} \E_n + p_0^{-} \overline{ \E_n}) + p_0^{+} (\overline{q_1^{-} \E_n + p_0^{-} \overline{ \E_n}} )\;.
\ee
The coefficients  $q_1^{+}$ and $p_0^{+}$ are defined by the formulas  \eqref{q_p_ceof_1},\eqref{q_p_ceof_2},\eqref{q_p_ceof_3}; they 
correspond to 
addition of   two solitons at $-\sigma + (n+1)L$ and $\sigma + (n+1)L$.  The coefficients $q_1^{-}$ and $p_0^{-}$ correspond to the addition of two solitons at $-\sigma - (n+1)L$ and $\sigma - (n+1)L$.

Assuming the convergence of our iterative procedure we define the limit
$$
\lim_{n \rightarrow \infty}  \Psi_n = \Psi_{\infty},
$$
which we naturally assume to be  the $\Psi$-function of the periodic analogue of Kerr solution.

The input parameters of the procedure are  $\sigma$, $L$  and $\alpha= p + iq$, while the remaining constant re-scaling of the Ernst potential  not yet fixed (see \eqref{scaling_E} and discussion at the end of \autoref{exact_formulas_coeff}). 

In the static case, the freedom in the choice of this multiplicative  constant is completely determined by the counter-terms $-\frac{2M}{nL}$ added in \eqref{U_PS} (at each step of addition of two solitons) to make the series convergent. 

Two regularization procedures of the Ernst potential lead to results which differ by a multiplicative real constant. In the non-linear stationary case we consider three different ways in which to fix this constant which are  convenient in various contexts. 

First, we study the convergence of the method by choosing the same counter-terms as in the static case (i.e., \eqref{U_PS}).

Once we have computed $\E_{n+1}$, \eqref{E_n+1_after_E_n}, we redefine $\E_{n}$ as
\be \label{regularization_1}
\E_{n} \mapsto \E_{n}e^{\frac{-4M}{nL}}, \quad n \geq 1
\ee
where $M = \frac{\sigma}{p}$ (we square the regularization term  since we add  four solitons at each  step). The overall constant is then fixed by taking the limit $n \rightarrow +\infty$.

Second, we compare the results of the iterative procedure with  the numerical solutions of \cite{Peraza:2022xic} in \autoref{comparison_sub_section}. In this case, we fix  the  area of the horizon $A$ to be $16 \pi$. Practically, we first use the previous regularization, and for the limiting value of $\E_\infty$ we scale it such that the area of the horizon is $16 \pi$.

Finally, when studying the asymptotic behaviour of solutions, in particular the existence domain in \autoref{sec_existence}, we regularize the sequence by fixing the value of $\E_{n}$ to be 1 at some reference point $(\rho_0 , z_0)$. Then, at each step , we  regularize as follows,
\be \label{regularization_2}
\E_{n} \mapsto \frac{\E_{n}}{\E_{n}(\rho_0 , z_0)}, \quad n \geq 1\;.
\ee
Here, to make the code more computationally efficient, we use fewer points in the grid, and choose the regularization \eqref{regularization_2} at $(L/2 , \rho_{MAX})$, which is  the farthest point in the numerical grid from the axis.

The solution generating method described above is very suitable for numerical analysis. In order to compute the value of $\E_n$ at $(\rho,z)$ for certain $n$, we only need the values of $\E_{n-1}$ at $(z,\rho)$, the constant $\alpha$ and the location of the new solitons which are placed at  $\pm \sigma \pm nL$. Moreover, the outcome at step $n$ does not depend 
 on the order in which the solitons are added.

We study three numerical  aspects of the solution generating method applied to periodic solutions. 

First, we study the convergence of the method, addressing the conjectured convergence of the sequence $\{\E_n\}$ proposed in \cite{Korotkin:1994cp}.  Here, we use regularization \eqref{regularization_1}.

Second, we compare the results of the iterative procedure with  the numerical solutions of \cite{Peraza:2022xic}. For that we use \eqref{regularization_1} and we scale the limiting solution $\E_\infty$ such that the area of the horizon equals to $16 \pi$.  
 
Finally, we study the asymptotic behaviour of the solutions and 
describe  the \textit{existence} domain in terms of the initial parameters $(\alpha , \sigma)$. Here, to make the code more computationally efficient, we use fewer points in the grid, and choose the regularization \eqref{regularization_2} at some point $(z_0 , \rho_0)$ which is  the farthest point in the numerical grid from the axis. 

\black

\subsection{Numerical Implementation} \label{sub_sec_num}

We implement the iterative construction on a numerical grid adapted to a finite region in which the Weyl coordinates range are as follows,
\[
(\rho,z) \in [\epsilon_\rho, \rhomax]\times [-L/2, L/2]\;,
\]
where $\epsilon_\rho > 0$ due to the possible singular behaviour of some functions at the axis. This domain represents a discretization of the fundamental domain of the periodic solution, and we refer to it as the \textit{central stripe} (due to its position in the $(\rho,z)$ plane).

For most of the numerical simulations, we use uniform grids, with $N_z$ points along $z$, semi-displaced with respect to the boundaries $z = \pm L/2$, and $N_\rho$ points along the $\rho$-direction:

\begin{equation}\label{grid_unif}\begin{split}
\rho_i &= \epsilon_\rho + \frac{\rho_{MAX}}{N_\rho - 1} i \hskip0.9cm  i = 0,...,N_\rho- 1,\\ 
z_j &= -\frac{L}{2} + \frac{L}{N_z} \Bigl( j + \frac{1}{2} \Bigr) , \quad j =
0,...,N_z-1
\end{split}
\end{equation}
where $\epsilon_\rho$ is a cut-off distance from the axis to avoid possible divergences. In most cases, $\epsilon_\rho \approx 10^{-3}$. Typical values of the numerical parameters are 
$$
N_z = 40,80, \quad N_\rho = 40,80, \quad  \rho_{MAX} = 5L , 10L , 40L
$$

Convergence and comparison with \cite{Peraza:2022xic} is performed using spectral and pseudo-spectral methods. We use a $N_\rho$-point Chebyshev grid to discretize $\rho$ and a uniform grid of $N_z$ points along $z$, semi-displaced with respect to the boundaries $z = \pm L/2$. 

\begin{equation}\label{grid}\begin{split}
\rho_i &= \epsilon_\rho + \frac{1}{2}\rhomax \left( 1 - \cos \Bigl( \frac{\pi}{N_\rho}i
\Bigr) \right) \quad i = 0,...,N_\rho,\\ 
z_j &= -\frac{L}{2} + \frac{L}{N_z} \Bigl( j + \frac{1}{2} \Bigr) , \quad j =
0,...,N_z-1.
\end{split}
\end{equation}

Observe that the axis $\{z =0\}$ is not included in the grid. The specific choice of grid  allow us to perform high precision derivatives and integrals. Along the $z$-direction, we approximate derivatives by the derivatives of the standard Fourier interpolation, while along the $\rho$-direction we approximate by derivatives of the polynomial interpolation.

To check different properties of the convergence of the sequence $\{\E_n\}$, we use an uniform grid along the $\rho$-direction, and the derivatives were approximated by the usual finite-difference 4th order formula for the derivative,
$$
f'(x_i) =  \frac{-f(x_{i+2}) + 8f(x_{i+1}) - 8f(x_{i-1}) + f(x_{i-2})}{12h} + O(h^4)\;.
$$

The locality and deterministic aspect of the computation implies the \textit{grid - independence} of the method we are using to obtain the Ernst potential. In other words, if we compute $\E$ on any two grids $\mathcal{G},\mathcal{G}'$ that coincide at a set of points $\mathcal{C} = \mathcal{G} \cap \mathcal{G}'$, then $\E \mid_{\mathcal{C}}$ is the same in both grids.

Nevertheless, any integration or differentiation that we compute from $\E$ can be understood as a functional $G_{\mathcal{G}}$ that depends on the particular grid we are using. Let $h$ be the smallness parameter of the numerical method representing consistently the linear problem, i.e. the discretization  parameter of the grid $\mathcal{G}$, and let $G[\E_n]^h$ be the \textit{numerical} computation of some functional on the solution we compute. Let $G[\E_n]$ be the value of the functional on the solution to the Ernst equation. Then, we can compute the order of convergence from $G[\E_n]^h$ to $G[\E_n]$ evaluating the quotient
$$
Q_{grid} = \frac{\norm{G[\E_n]^{h} - G[\E_n]^{h/2}}}{\norm{ G[ \E_n]^{h/2} - G[\E_n]^{h/4}} },
$$
where the norm is the discrete version of $L^2$ on the numerical domain.

To compute the convergence with respect to the number of solitons added, i.e. the parameter $n$, we study the uniform convergence using the discrete version of the $L^2$ norm. First, we look at the relative velocity of convergence, 
\be \label{rel_rate_of_convergence}
v_c =  \frac{\norm{\E_{n+2} - \E_{n+1}}}{\norm{\E_{n+1} - \E_{n}} }\;.
\ee
This gives us an estimate of the convergence velocity.

Next, we compute the (empirical) order of convergence and rate of convergence via the formulas
\be \label{order_of_convergence}
\bar{Q} = \frac{\ln \left( \frac{\norm{\E_{n+3} - \E_{n+2}}}{\norm{\E_{n+2} - \E_{n+1}} } \right)}{\ln \left( \frac{\norm{\E_{n+2} - \E_{n+1}}}{\norm{\E_{n+1} - \E_{n}} } \right)} ,
\ee
\be \label{rate_of_convergence}
\mu =  \frac{\norm{\E_{n+1} - \E_{\infty}}}{\norm{\E_{n} - \E_{\infty}}^{q} },
\ee
where in the last equation we use the empirical order of convergence $\bar{Q}$ in the exponent, and where $\E_{\infty}$ is regarded as the \textit{numerical solution} to the periodic setup, in the sense that it corresponds to some large stopping value of $n$ we chose. We usually take $\E_{\infty} = \E_{4000}$, based on the results we obtained for several runs.

We review the iterative approach step-by-step.

\begin{enumerate}
\item Fix the number of solitons to be added, $N_s$. For convenience, we use $N_s$ of the form $4k + 2$, such that we have an even number of black hole on each side of the central stripe.

\item Initialize the matrix $\Psi_0$ corresponding to an empty background, the locations of the solitons and their respective constants with values $\alpha$ and $-\alpha^{-1}$.

\item Choose the order in which the solitons are added. As a rule of thumb, to avoid zeros from the Ernst potential  in a neighborhood of the horizon (cf. \autoref{erg_sub}), the last soliton to be added is the one corresponding to the central stripe, while the rest are added in any order. 

\item On each step $1 \leq n \leq N_s/2$, we solve $T_n$ via \eqref{q_p_ceof_1},\eqref{q_p_ceof_2},\eqref{q_p_ceof_3}, with the values for $\beta$'s are computed from \eqref{leading_coef_condition} using $\Psi_n$ as the background solution. We obtain $T_n$ directly via \eqref{E_n+1_after_E_n}, and the regularization procedure (either \eqref{regularization_1} or \eqref{regularization_2}).

\item At the final step, we obtain the value of the Ernst potential $\E_{(N_s - 2)/4}$.

\end{enumerate}

Observe that the computational time of the whole construction behaves as  $O(N_z N_\rho N^2_s/2)$, given that on each loop one  has to actualize the values of the dressings for $\alpha$'s.

Finally, to measure the error within which equation \eqref{Ernst_equation} is satisfied, we compute the quotient between the $L_2$ norms of the operator 
$$
\mathfrak{E}(\E):=(\E + \Eb) (\E_{zz} + \frac{1}{\rho}\E_\rho + \E_{\rho \rho}) - 2 (\E_{z}^2 + \E_\rho^2)
$$
and the function $\E$, that is, the relative error in the Ernst equation,
\be  \label{error_relative_solution}
\epsilon_\mathfrak{E} := \frac{ \| \mathfrak{E}(\E)\|}{\|\E \|}\;.
\ee

\section{Numerical results and Analysis} \label{sec_results}

In this section we present the convergence analysis of multi-soliton approach.

\begin{figure}[h!]
\centering
\includegraphics[scale=0.3]{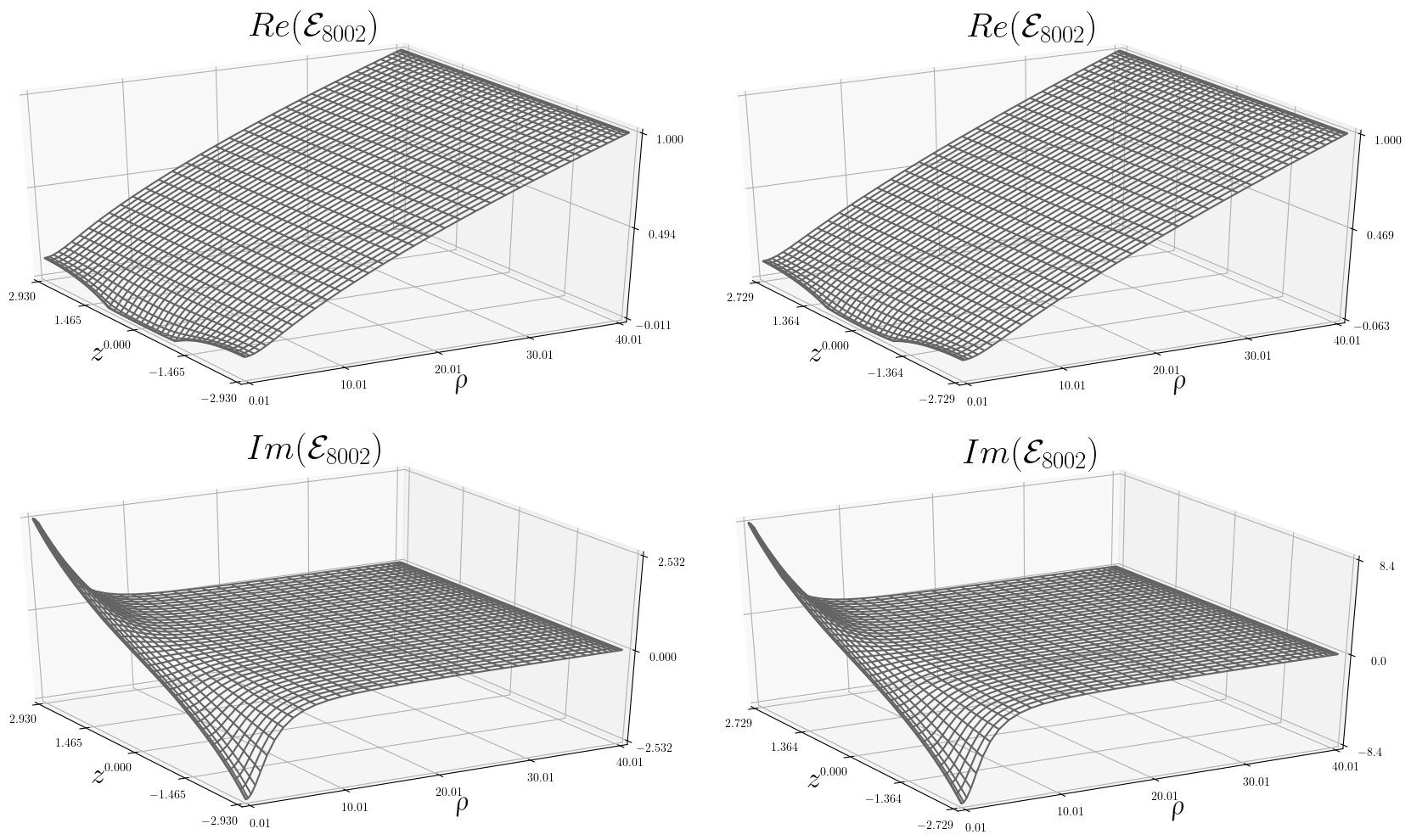}
\caption{Plot of the real part of the Ernst potential, $\re\E$, using the iterative method for 8002 solitons. The values of  parameters $(\sigma, \alpha)$ are $(0.9768... , 0.9692...)$(left) and $(0.9095... , 0.8824...)$(right).} \label{examples}
\end{figure}

In figure \ref{examples} we present two typical examples of solutions for $(\sigma , \alpha)$ corresponding to
$$
(\sigma_1 , \alpha_1) = (0.9768... , 0.9692...),
$$
and 
$$
(\sigma_2 , \alpha_2) = (0.9095... , 0.8824...).
$$
The angular momentum corresponding to the pair $(\sigma_1 , \alpha_1)$ is equal to $ J= 0.25$, and for the pair $(\sigma_2 , \alpha_2)$ we have  $J=0.5$ These parameters are chosen such that the horizon area of the corresponding asymptotically flat black hole equals $A = 16 \pi$, and therefore the angular momentum is bounded from above, $0 \leq |J| \leq 2$ (cf. \cite{Peraza:2022xic}).

The solutions are  normalized such that the maximum of the modulus of $\E$ on the numerical grid equals to  1.

\subsection{Convergence analysis of the series}

\subsubsection{Error in the numerical solution of Ernst equation}

First, we check that the iterative method indeed  produces a solution of the Ernst equation \eqref{Ernst_equation}. We use the relative error \eqref{error_relative_solution} to estimate how far from an actual solution we are at each step.

\begin{figure}[h!]
\centering
\includegraphics[scale=0.5]{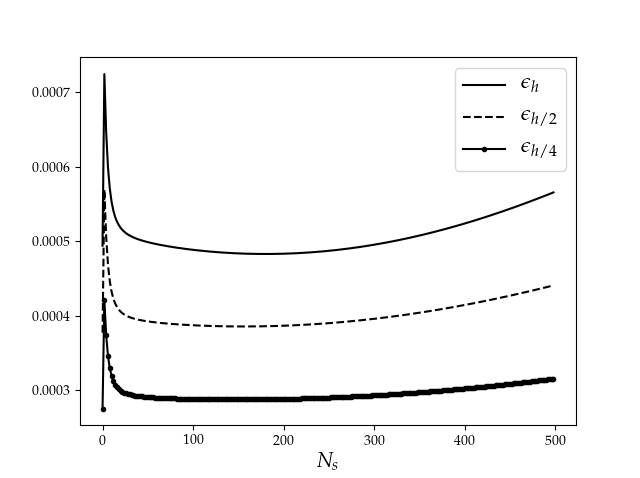}
\caption{Relative error \eqref{error_relative_solution} for the first 500 soliton pairs added.} 
\label{error_figure}
\end{figure}

Of course, the error is grid dependent, since the derivatives with respect to the variables $z$ and $\rho$ are computed numerically. In \autoref{error_figure} we show the relative error for three runs with uniform grids with sizes $(N_{\rho} , N_z) = (20,20) , (40,40) , (80,80)$. The grid size, denoted by $h$, decreases by half at each step. As it can be seen, the error decreases as the grid size decreases. 

Computing the $Q$-factor for the error in the Ernst equation, we obtain value close to 1,
$$
Q_{\epsilon_{\mathfrak{E}}} \approx 1.0035,
$$
which implies a slightly faster rate of convergence than that of linear convergence for $\epsilon_{\mathfrak{E}}$. Then, we can assume that the sequence of numerical solutions stays within the class of solutions to the Ernst equations, with tolerance of $\approx 0.1\%$.

\subsubsection{Relative error in the sequence}

We provide strong numerical evidence for the convergence of the sequence $\{\E_n \}$ by computing the logarithmic error with respect to the numerical solution (which we take for a large $N_s \approx 10^4$),
$$
\ln \varepsilon = \ln \left( \frac{ \norm{\E_{n} - \E_{\infty}} }{ \norm{ \E_{\infty} }} \right)
$$
and the relative error between consecutive terms,
$$
\ln \hat{\varepsilon} = \ln \left( \frac{ \norm{\E_{n+1} - \E_{n}} }{ \norm{ \E_{n} }} \right)
$$

\begin{figure}[h!]
\centering
\includegraphics[scale=0.4]{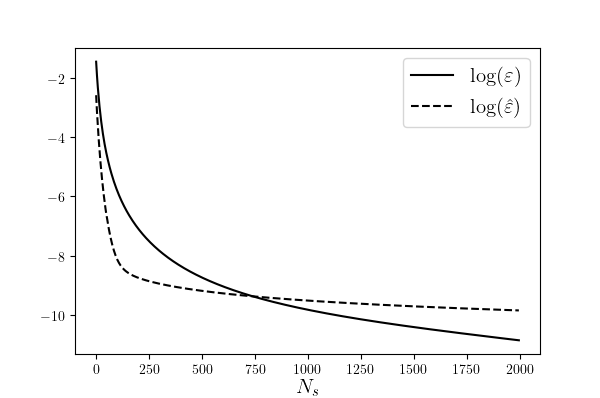}
\includegraphics[scale=0.4]{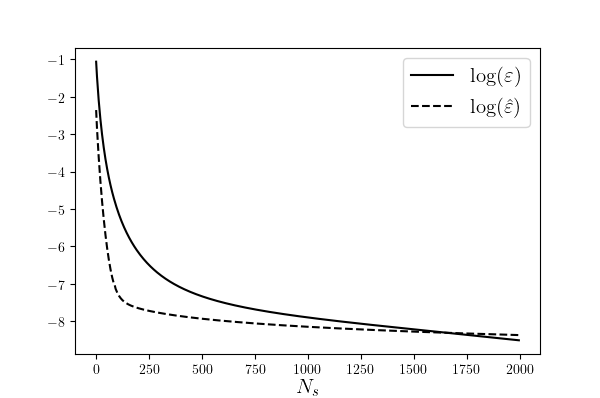}
\includegraphics[scale=0.4]{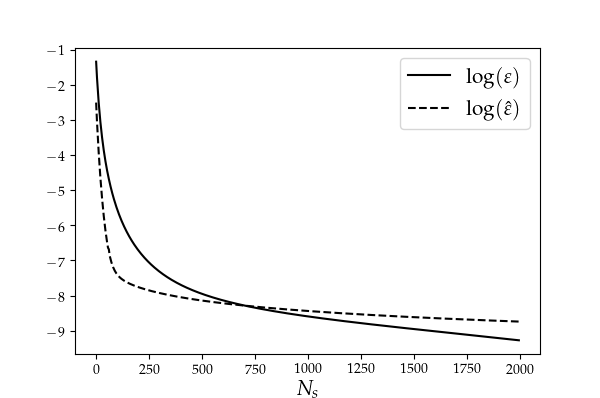}
\caption{Relative error and consecutive relative error of the sequence $\{\E_n \}_{n\geq 0}$. From left to right and from top to bottom the values $(\sigma/L , J)$ are $(0.125 , 0.25) $, $(0.167 , 0.25)$ and $(0.125, 0.5)$.}
\end{figure}

\subsubsection{Order and rate of convergence}

To study the order and rate of convergence of the sequence $\E_n$ we use formulas \eqref{rel_rate_of_convergence}, \eqref{order_of_convergence} and \eqref{rate_of_convergence}. Below we give three values of orders and rates of convergence for different selections of parameters, and for a fixed grid size of $(40,40)$.
\begin{table}[h!]
\captionsetup{margin=1cm}
\begin{center}
\begin{tabular}{c|c|c|c|c}
$\sigma/L$ & $J$ & $\bar{Q}$ & $\mu$ & $v_c$ \\ 
\hline 
$\frac{1}{6}$ & 0.25 & 0.99779 & 0.99790 & 0.99931 \\ 
$\frac{1}{8}$ & 0.25 & 0.99778 & 0.99774 & 0.99892 \\ 
$\frac{1}{8}$ & 0.5 & 0.99778 & 0.99779 & 0.99905 
\end{tabular} 
\end{center}
\caption{\label{table_convergence} Convergence order, rates, and relative velocity for different values of $\sigma/L$ and $J$.}
\end{table}

Observe that we obtain consistently values of the order and rate of convergence close to 1. That means a logarithmic convergence of the sequence.

\vspace{0.5cm}

\textbf{Comparison with the static case $J=0$}

\vspace{0.5cm}

\begin{figure}[h!]
\centering
\includegraphics[scale=0.5]{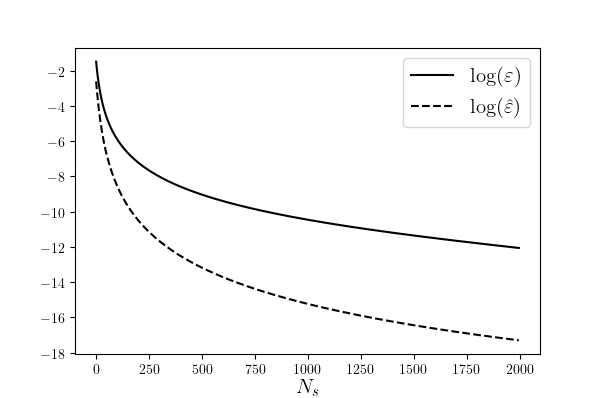}
\caption{\label{MKN_errors} Plots of logarithmic errors for Periodic Schwarzschild case, in terms of the number of solitons.}
\end{figure}

We can set $\alpha \equiv 1$ and compute numerically the solution to the (linear) Ernst equation, now simply by adding up the solitons. Then we arrive to the periodic analogue of Schwarzschild. The order and rate of convergence in this simple  case can be compared with the ones corresponding to the non-linear case discussed above, see \autoref{MKN_errors}.

\begin{table}[h!]
\captionsetup{margin=1cm}
\begin{center}
\begin{tabular}{c|c|c|c|c}
$\sigma/L$ & $J$ & $\bar{Q}$ & $\mu$ & $v_c$ \\ 
\hline 
$\frac{1}{6}$ & 0 & 0.99963 & 0.99812 & 0.99120   \\ 
$\frac{1}{8}$ & 0 & 0.99756 & 0.99443 & 0.99336
\end{tabular} 
\end{center}
\caption{\label{table_convergence} Convergence order, rates, and relative velocity for $\sigma/L = 1/6, 1/8$ and $J=0$.}
\end{table}

\subsection{Comparison with the Harmonic Map Heat Flow Method} \label{comparison_sub_section}

The numerical approach to the solution of the periodic problem for the  Ernst equation adopted in  \cite{Peraza:2022xic} was based on the harmonic map heat flow. 
 The initial value problem used in \cite{Peraza:2022xic} is defined by three parameter, $(\sigma/L,A,J)$, and  a multiplicative constant for the conformal factor. Then, a system of coupled partial differential equations is solved with suitable boundary conditions.

In order to compare the numerical methods applied to the same initial value problem, we have to chose appropriate values for $\alpha$ and the normalization constant for the Ernst potential. 
In this paper the solution is defined by  three non-physical parameters $(\sigma, p,L)$. The map from these parameters to the set of parameters $(\sigma/L,A,J)$ is by no means trivial.
In particular, the condition  $A = 16 \pi$ translates into fixing of the value of $\sigma$ via equation (e.g. \cite{stephani_kramer_maccallum_hoenselaers_herlt_2003})
\be 
\sigma = \sqrt{\frac{A}{16\pi}} \frac{ 1 - \left( 8\pi J/A\right)^2 }{\sqrt{1 + \left( 8\pi J/A \right)^2}}. 
\ee

In our numerical method, we do not have an a priori knowledge of the angular momentum of the final solution. It has to be computed once we have the solution, via \eqref{Komar_ang_mom}. Nevertheless, we can associate unambiguously a value of $p_0$ from the value of $J$, via the formulas for the asymptotically flat Kerr black hole. Indeed, let $M_{Kerr}(\sigma , J)$ be the mass of an asymptotically flat Kerr black hole which corresponds to the same value $J$ of the angular momentum as the periodic Kerr solution. Then $M_{Kerr}$  can be found from the equation 
$$
M^2_{Kerr} = \sigma^2  + \frac{J^2}{M^2_{Kerr}}\;.
$$
Then we define
$$
p_0 = \frac{\sigma}{M_{Kerr}}.
$$
We can use the value $p_0$ as a first approximation  for the  value of the constant $p$ used as an input to the periodic Kerr solution. 
Then one can use an iterative bi-partition method to get the  value of $p$ which provides the correspondence between 
our current construction and  the construction of \cite{Peraza:2022xic}.

We find that the quotient of $\Re\E$ of the  periodic black hole  solutions obtained in  \cite{Peraza:2022xic}  and in this paper (for the same set of parameters
$(\sigma/L,A,J)$) to be  {\it almost a constant}, within $\approx 0.2 \%$ in relative error. See below \autoref{table_comparison}.

\begin{table}[h!]
\begin{center}
{\footnotesize
\begin{tabular}{cccccc}
$(\sigma,p,L)$ & Relative error (mod $cnt.$) &  $\kappa(ISM)$ &  $\kappa(HMHF)$ \\
\hline
(0.9768, 0.9909, 8.8798) & $5.1 \times 10^{-4}$  & $4.5520 \times 10^{-1}$ & $4.5477 \times 10^{-1}$ \\
(0.9768, 0.9909, 6.9770) & $8.3 \times 10^{-4}$ & $5.7962 \times 10^{-1}$ & $5.8017 \times 10^{-1}$ \\
(0.9768, 0.9909, 5.7457) &  $9.6 \times 10^{-4}$ & $7.1007 \times 10^{-1}$ & $7.0768 \times 10^{-1}$ \\
(0.9768, 0.9909, 4.8839) &  $1.5 \times 10^{-3}$ & $8.3288 \times 10^{-1}$ & $8.3977 \times 10^{-1}$ \\
\hline 
(0.9095, 0.9528, 8.2683) & $8.7 \times 10^{-4}$ & $5.0058 \times 10^{-1}$ & $5.0278 \times 10^{-1}$ \\
(0.9095, 0.9528, 6.4965) & $9.6 \times 10^{-4}$ & $6.5129 \times 10^{-1}$ & $6.4781 \times 10^{-1}$ \\
(0.9095, 0.9528, 5.3501) & $1.2 \times 10^{-3}$ & $8.0973 \times 10^{-1}$ & $8.0632 \times 10^{-1}$ \\
(0.9095, 0.9528, 4.5475) & $1.7 \times 10^{-3}$ & $9.9085 \times 10^{-1}$ & $1.0003$ \\
\end{tabular}
}
\caption{\label{table_comparison} Relevant quantities computed for the
solutions in the series with $J=1/4$ (first four lines) and $J=1/2$ (second four lines). The values of the Kasner exponent shown here were computed using the inverse scattering method $(ISM)$ and the harmonic map heat flow $(HMHF)$ from \cite{Peraza:2022xic}.} 
\end{center}
\end{table}

%

\subsection{Metric coefficients and Ergosphere}

In this section we show how to compute the rest of the metric coefficients, $F$ and $k$, to completely characterize the metric in \eqref{ernst_metric_form}. 
We use two different approaches. First, at each loop of the run, we can use formulas \eqref{F_from_F_0} and \eqref{k_from_k_0} to obtain the exact formula for $F$ and $k$ recursively. On the other hand, we can integrate equations \eqref{conformal_factor} directly using the numerical values of the Ernst potential. 

\begin{figure}[h!]
\centering
\includegraphics[scale=0.3]{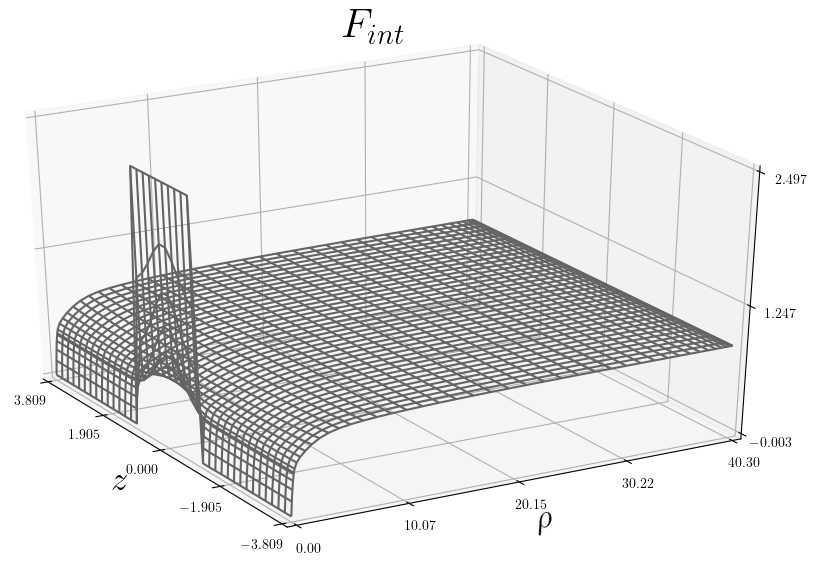}
\includegraphics[scale=0.3]{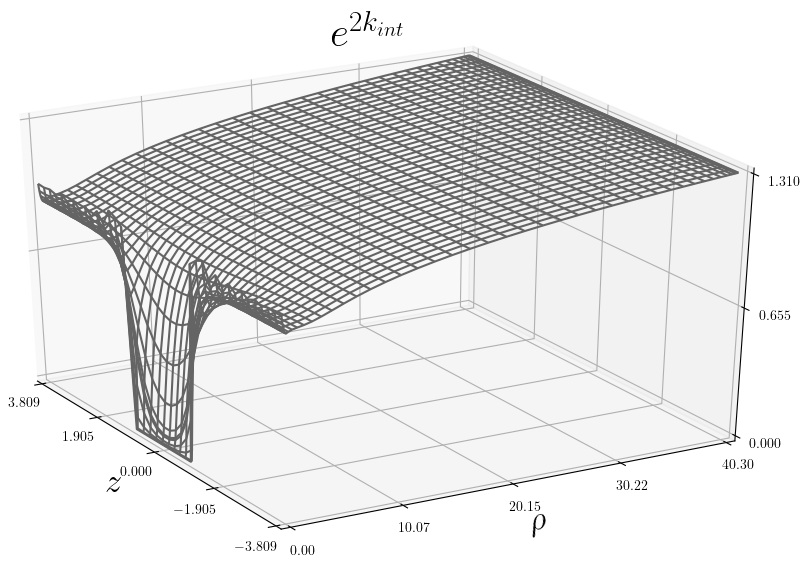}
\caption{Metric coefficients $F$ (top) and $e^{2k}$ (bottom), both computed via integration of equations \eqref{conformal_factor}. } \label{metric_coeff}
\end{figure}

These methods are compared  for parameters, $\sigma/L = 0.125$, $p= 0.9909$. We take $N_s\approx 10^4$. 
The computation of the quadratures is very sensitive to noise close to the axis, due to the singular behaviour of the coefficients (cf. \eqref{conformal_factor}). To avoid this, we carry the integration with grid points up to $\rho = 0.3$, and for points with $\rho$ coordinate in the interval $(0,0.3)$ we use interpolation (standard cubic interpolation). In \autoref{metric_coeff} we show the result of the integration.

\begin{figure}[h!]
\centering
\includegraphics[scale=0.45]{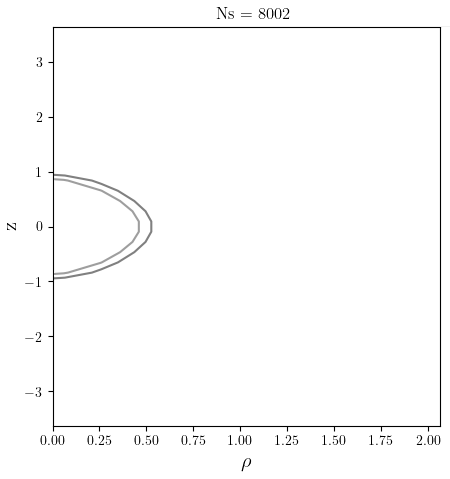}
\caption{Plot of the ergosphere, $\text{Re}\E=0$, for  $\sigma/L = 0.125$, $J= 0.25$ and $N_s = 8002$. As a reference, in light colour we draw the ergosphere for the asymptotically flat Kerr solution with the same parameters.} \label{ergosphere}
\end{figure}

In \autoref{ergosphere} we show a typical example of the ergosphere (black), with the asymptotically flat (light gray) as a reference.

\subsection{Numerical evaluation of Kasner exponent}

The Kasner exponent can be computed by analyzing the asymptotic behaviour of   $\text{Re}\E$, since asymptotically we have
\be 
f \rightarrow C \rho^\Kasner \quad (\rho \rightarrow + \infty)
\ee

\begin{figure}[h!]
\centering
\includegraphics[scale=0.6]{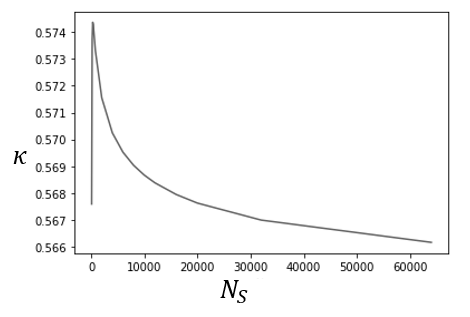}
\caption{Example of convergence of Kasner exponent as the  number of solitons $N_s$ tends to infinity.} \label{kasner_exp}
\end{figure}

By adjusting the far end of the numerical interval in $\rho$ variable (e.g. the largest ten point), we can obtain the coefficient $\Kasner$. In terms of the number of solitons, we see in \autoref{kasner_exp} that indeed the value converges as the number of solitons tends to infinity.

\begin{figure}[h!]
\centering
\includegraphics[scale=0.5]{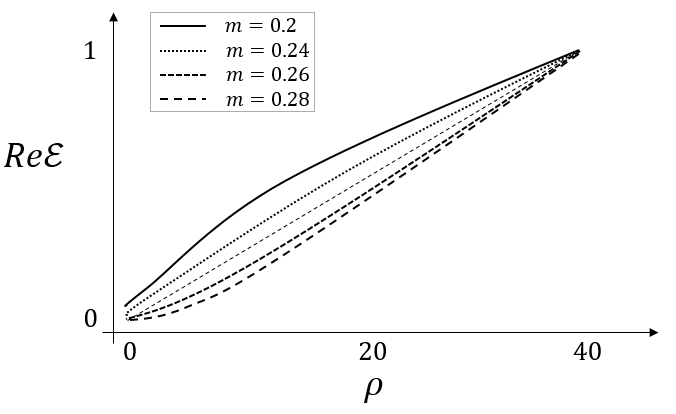}
\caption{Behaviour of the $z-$average of $\text{Re}\E$ for four solutions with $m$ near $0.25$. The dashed line indicates a linear growth. All solutions were normalized by $\re \E \mid_{\rho_{max}} = 1$ so that the change in concavity is easily visible.} \label{asymptotic}
\end{figure}

In \autoref{table_kasner_exp} we show some typical values for the Kasner exponent in a series of numerical solutions computed with increasing values of $m$. As can be clearly seen from the numerical solutions, when $4m > L$ the Kasner exponent $\Kasner$ is greater than 1, which implies that $f^{-1} \rho^2$ growths sub-linearly. In \autoref{asymptotic}, with the dotted line being a linear growth, we show the different behaviour of four solutions near $m= 0.25$. 

\begin{table}[h!]
\begin{center}

\begin{tabular}{cccccc}
$L$ & $m$ & Angular velocity &  $\kappa$ (from $V$) \\
\hline
1 & 0.28 & $2.2380 \times 10^{-1}$ & 1.1308 \\
1 & 0.26 & $1.8102 \times 10^{-1}$ & 1.0501 \\
1 & 0.24 & $1.3764 \times 10^{-1}$ & 0.9696 \\
1 & 0.20 & $9.6651 \times 10^{-2}$ & 0.8090 \\
1 & 0.18 & $7.8501 \times 10^{-2}$ & 0.7290 \\
1 & 0.16 & $7.1450 \times 10^{-2}$ & 0.6490 \\
1 & 0.14 & $6.6909 \times 10^{-2}$ & 0.5683 \\
\end{tabular}

\caption{\label{table_kasner_exp} Relevant quantities computed for the
solutions in the series with $J=1/4$.} 
\end{center}
\end{table}

\section{Numerical description of existence domain } \label{sec_existence}

The numerical implementation presented in \autoref{sec_setup} can be used to compute the existence domain of the solution in the $(p,\sigma)$-plane, by fixing the periodic length to a certain value, say $L=4$.

We simplify the grid in the $(\rho,z)$-plane, and analyze the  limit $N_s\to\infty$ on the line $z=-L/2$. On this line, we take a sequence of equidistant points far away from the axis. Given the value of $L$, we compute the value of the Kasner exponent by fitting points located between $5L$ and $10L$.

Assuming a fall-off for the real part of the Ernst potential as follows, 
$$
f \approx \rho^\Kasner + O(\ln \rho), \quad \Kasner < 1\;,
$$
the $\rho$-derivative of the logarithm of $f$ decays faster than
$$
\partial_\rho \ln f \approx \frac{\Kasner }{\rho} + O(1 /(\rho \ln(\rho)) )\;.
$$
Then, we can extract $\kappa$ by fitting the \textit{numerical quantity}
$$
\rho  \partial_\rho \ln f
$$
with a function of the form $\Kasner + \frac{b}{\ln \rho}$.

%

\vspace{0.2cm}

\textbf{ Remarks about the existence graphs.} We can compare the deviation of the mass value with respect to the asymptotically flat case. Since we are fixing the area of the horizon to be that of Kerr solution for the same $\sigma$ and $J$,  this comparison is valid in the whole range of parameters.

In view of the numerical results, we have that $\Kasner \geq \frac{4\sigma}{p L}$. Below we give a physical interpretation of this phenomena in terms of the rotational energy of the horizons. Assume that
\be \label{kappa_less}
\Kasner < \frac{4\sigma}{p L}.
\ee
By Smarr identity, we have
$$
\sigma + 2 \Omega J = \frac{\Kasner L}{4} < \frac{\sigma}{p} \;.
$$
Consider a periodic solution with the \textit{same} angular momentum as the isolated asymptotically flat Kerr solution with parameters $(p,\sigma)$. Then, since $ J_{Kerr} = \frac{\sigma^2}{p^2} \sqrt{1 - p^2}$, we have
$$ 
\Omega < \frac{p}{2 \sigma} \sqrt{\frac{1-p}{1+p}}  =\Omega_{Kerr} \;,
$$
where the right hand side is the asymptotically flat value of the angular velocity for a Kerr black hole with parameters $(p,\sigma)$. Then we arrive at the conclusion that, if \eqref{kappa_less} is valid, then for the same energy input the periodic analog of the Kerr black hole rotates with less angular velocity than its asymptotically flat counter-part. This is, of course, in contradiction with the physical intuition that two identical rotating bodies rotate \textit{faster} than a single isolated body due to frame dragging, given that the input energy is the same.



\subsection{Non-physical vs. physical sets of  parameters}

To construct a solution via the method presented in this work, we need to specify a triplet of  parameters $(\sigma, p, L)$ which are convenient 
from the numerical point of view but don't have a direct physical interpretation. The   parameters which have the direct physical meaning are $(D,J,A)$, where 
$$
D= \int_{\sigma}^{L-\sigma} \frac{e^k}{f^{1/2}} dz
$$ 
is the distance between two subsequent horizons, $J$ is the angular momentum of the periodic Kerr solution defined as an appropriate Komar integral around the event horizon (cf. \eqref{Komar_ang_mom}) and $A$ is the area of each horizon. We also define the 
meridian distance between South and North pole,
$$ 
\ell = \int_{-\sigma}^{\sigma} \frac{e^k}{f^{1/2}} dz\;.
$$
We construct families of solutions in terms of  the  variables $(\sigma, p, L)$, and then compute the dependence of the solutions on the physical variables.

\subsubsection{Variables $(\sigma, p ,L)$}

Fixing $L=4$, we compute the Kasner exponent of each configuration, as discussed above. Non-singularity of the solution in the full range of coordinates outside of the event horizon  is determined via the condition $\Kasner < 1$. 

\begin{figure}[h!]
\centering
\includegraphics[scale=0.5]{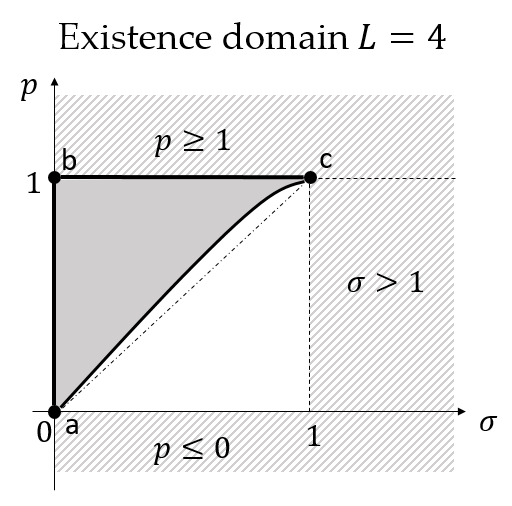}
\caption{Existence domain in terms of  parameters $(\sigma,p)$ and $L = 4$. In the light-grey stripped regions the non-existence of non-singular solution was proven theoretically. In solid grey region the numeric analysis suggests existence. In the white area the non-existence of non-singular solution.}
\label{existence_p_s}
\end{figure}

In \autoref{existence_p_s} we show the results obtained. The light-grey stripped regions indicate established regions of non-existence, while the grey solid region indicates the existence region for periodic Kerr. Analogous graphs can be done for different values of $L$, all presenting the same qualitative behavior of the non-existence/existence transition curve.

As a reference, segment lines joining the points a, b and c in the figure indicate: (a-b) asymptotically flat solutions, and (b-c) periodic Schwarschild solutions with $\sigma<1$.

From this picture alone, we cannot discard periodic extremal objects, since the line $p=0$ only contains solutions if $\sigma=0$. In other words, the point (a) contains the asymptotically flat extremal Kerr solution, but in principle it also contains periodic extremal horizons. We are going to study this in detail once we proceed to compute the physical parameters of the solutions.

\begin{figure}[h!]
\centering
\includegraphics[scale=0.4]{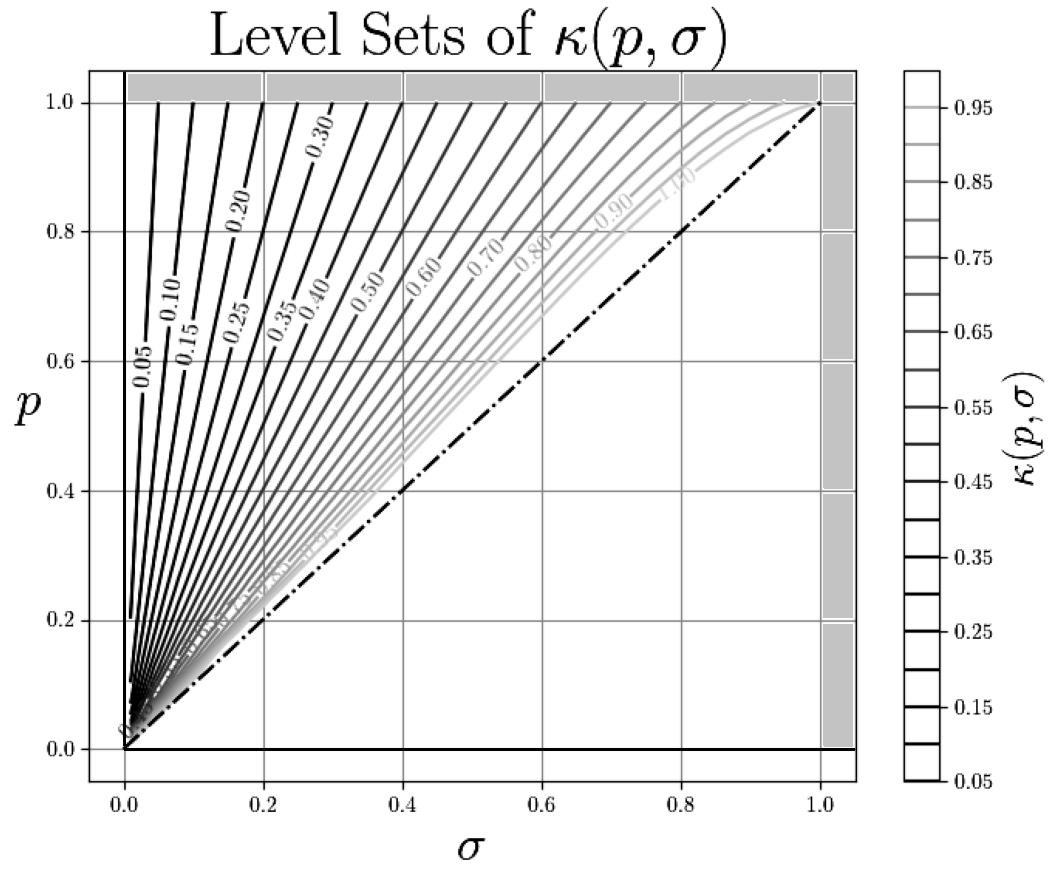}
\caption{Level curves for the Kasner exponent $\Kasner$ as function of  $p$ and $\sigma$ for $L=4$.}
\label{level_sets_Kasner}
\end{figure}

We can determine the level curves of the Kasner exponenet $\Kasner(\sigma,p)$ for given $L$, via standard bisection methods; they are shown in  \autoref{level_sets_Kasner}.

Observe that the level sets are equidistant at the line $p=1$. This is in correspondence with the analytic expression (\ref{KsL}) for the Kasner exponent of  periodic Schwazrschild solutions in terms of  $\sigma$ and $L$  \cite{Korotkin1994}, 
which in the case $L=4$ we are considering here gives  $\Kasner(\sigma ,4) = \sigma$.

\begin{figure}[h!]
\centering
\includegraphics[scale=0.45]{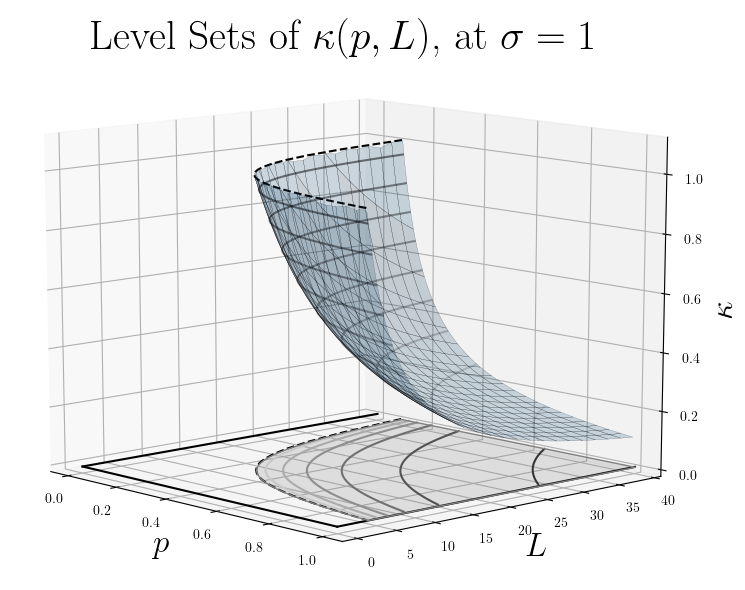}
\caption{Kasner exponent as function of $p$ and $L$ for fixed $\sigma =1$. The existence domain of the periodic Kerr  $\Kasner < 1$ is shown in grey. The dashed curve  $\Kasner = 1$ in the $(p,L)$-plane is the boundary of the existence domain. In solid lines we draw the level curves of $\Kasner(p,L)$. }
\label{Level_Sets_Kasner_p_L}
\end{figure}

Alternatively, one  can  fix $\sigma$ (we put $ \sigma= 1$) to describe the  region of existence in the $(p,L)$-plane together with the graph of $\Kasner$ and the level curves. The result is shown in  \autoref{Level_Sets_Kasner_p_L}.

\subsubsection{Variables $(\sigma /L , J ,A)$}

\begin{figure}[h!]
\centering
\includegraphics[scale=0.45]{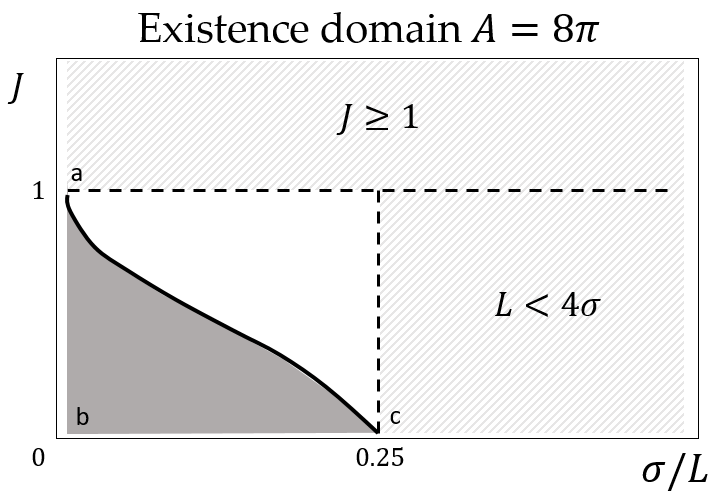}
\caption{Existence domain in $(J,\sigma/L)$ parameters for  fixed $A = 8\pi$. Regions shaded in light-grey: theoretical non-existence. Solid grey region: numerical existence. White area: numerical non-existence.}
\label{Existence_domain_J_sigma}
\end{figure}

To describe the asymptotically flat limit of the periodic Kerr  it is convenient to work in terms of variables $(\sigma /L , J ,A)$ and fix the area $A = 8\pi$. If we let the quotient $\sigma/L $ tend to $0$, then the periodic solutions approach the asymptotically flat solution corresponds, as it is shown in \autoref{Existence_domain_J_sigma}.

Observe that the values of the angular momentum are below the extremal limit $J=1$ \cite{Dain:2011pi}. We see no numerical evidence that supports the existence of periodic analogues of extremal Kerr solutions.

\subsubsection{Variables $(D , J ,A)$}

We compute $\Kasner$ as function of physical  $(D , J ,A)$ parameters  {\it  a posteriori}, meaning that we first find $\Kasner$ in terms of $(\sigma, p,L)$ and  also compute  physical parameters $(D , J ,A)$ in terms of $(\sigma, p,L)$. Then the change of variables produces $\Kasner(D , J ,A)$. In general, computing $\Kasner$ for fixed the angular momentum or the fixed area is straightforwards. 

\begin{figure}[h!]
\centering
\includegraphics[scale=0.45]{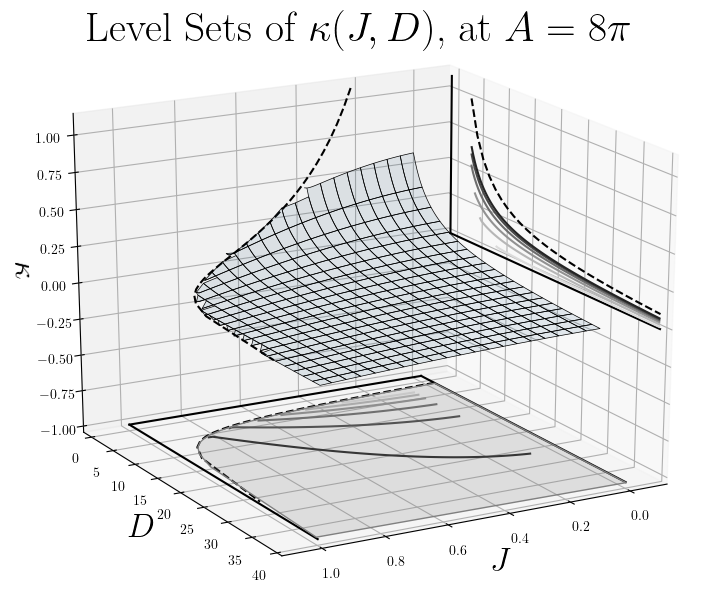}
\caption{Existence domain in $(J,D)$ parameters for fixed $A = 8\pi$. The existence domain $\Kasner < 1$ is shown in grey. The dashed line in the $(J,D)$-plane is  the level set $\Kasner = 1$. Solid lines are the level sets of $\Kasner(J,D)$. }
\label{Level_Sets_Kasner_J_D}
\end{figure}

The computation of $K$ for fixed distance between the horizons is less direct, since one needs to use a large set of solutions to obtain the desired value of $D$, via bisection method in the original variables (if one wants  a specific value for $D$) or the interpolation method (if only a  qualitative description is required).

In \autoref{Level_Sets_Kasner_J_D}, we show the plot of $\Kasner(J,D)$ for fixed area $A = 8 \pi$. This graph shows that non-existence/existence curve is \textit{very} sensitive to the values of $D$ and $J$ in a neighborhood of the curve $\Kasner = 1$.

\subsection{Critical distance}

Now we can compute the  critical physical distance between horizons, below which no periodic Kerr solution can exist.

Indeed, in \cite{Peraza:2024uto} it is shown that the analytic result of non-existence below $L = 4\sigma $ is equivalent to a statement between physical parameters. Namely, let $(\sigma , L)$ determine a periodic Schwarzschild solution, and let $D$ be the physical distance between the horizons and $A$ the area of each horizon. Then, if \footnote{We use the quantity $\sqrt{A/\pi}$ since, geometrically, it corresponds to the diameter of a sphere in Euclidean space.}
\be \label{bound_A_D_1}
\sqrt{\frac{A}{\pi}} \geq \frac{12}{\sqrt{\pi}} D \approx (6.7702...)\times D,
\ee
or
$$
A\geq 144 D^2,
$$
the solution does not admit an axisymmetric and stationary perturbation with non-zero angular momentum. This bound can be improved by using properties of Gamma functions \cite{Peraza:2023boa}, 

\be \label{bound_A_D_2}
\sqrt{\frac{A}{\pi}} \geq \frac{\sqrt{72}}{\Gamma_{min} \pi^{3/2}} D \approx (1.7316...)\times D\;,
\ee
where $\Gamma_{min} \approx 0.88560...$ is the minimum of the Gamma function on the interval $(0,2)$.

\begin{figure}[h!]
\centering
\includegraphics[scale=0.45]{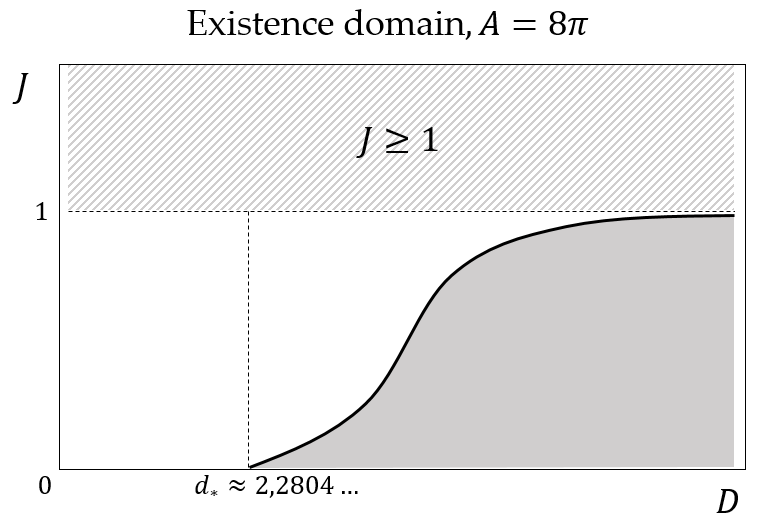}
\caption{Example of existence domain in terms of $(J,D)$, for $A=8\pi$. The hyperextremal region is shown in light-grey. The existence region is shown in grey. The value of the critical distance below which no static solution can be rotated is approximately $d_* = 2.2804...$}

We study the critical distance first by fixing the area $A$ and finding the non-existence/existence curve via the bisection method. In \autoref{Critical_value_Physical_param} we show a typical example, with $A = 8 \pi$

\label{Critical_value_Physical_param}
\end{figure}

\begin{figure}[h!]
\centering
\includegraphics[scale=0.35]{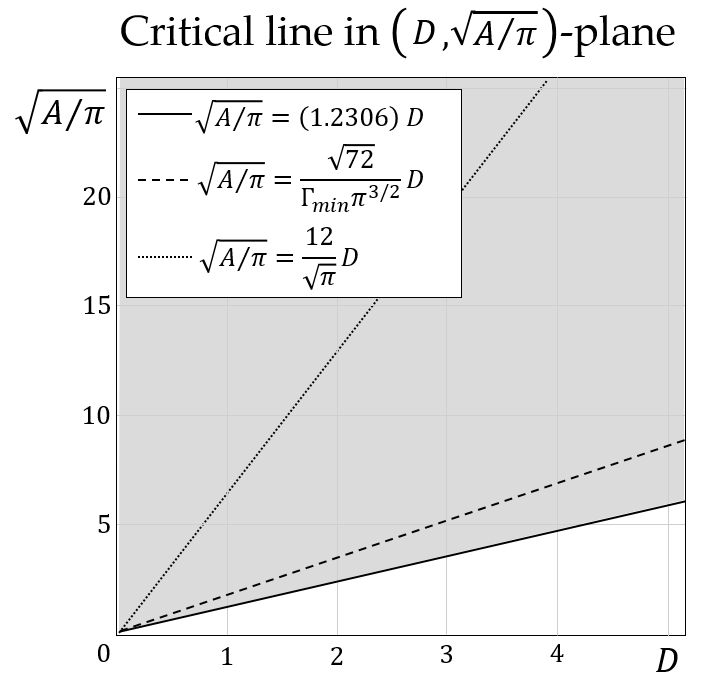}
\caption{The grey domain in the $(D, \sqrt{A/\pi})$ contains the range of parameters where a static solution can not be put in rotation obtained numerically. The  theoretical bounds, \eqref{bound_A_D_1} and \eqref{bound_A_D_2} are shown in dotted and dashed lines, respectively.}
\label{Critical_value_line}
\end{figure}

We can repeat the above procedure  for several values of the area, and fit to a linear model using ordinary least squares method (with an $r^2$ score of $0.95$). In \autoref{Critical_value_line} we show the non-existence/existence curve in a continuous line, while bound \eqref{bound_A_D_1} is shown in dotted line, and bound \eqref{bound_A_D_2} is shown in dashed line. 

\subsection{Ergosphere} \label{sub_sec_ergosphere}

The ergosphere, determined by the equation $\re\, \E = 0$, can be plotted straightforwardly form the calculations of the Ernst potential. In view of the results of  \cite{Peraza:2022xic}, and the results of ergosphere behaviour for black hole binaries \cite{Costa2009}, we expect that the ergospheres of adjacent rotating black holes tend to merge, since their angular momentums are aligned.

\begin{figure}[h!]
\centering
\includegraphics[scale=0.4]{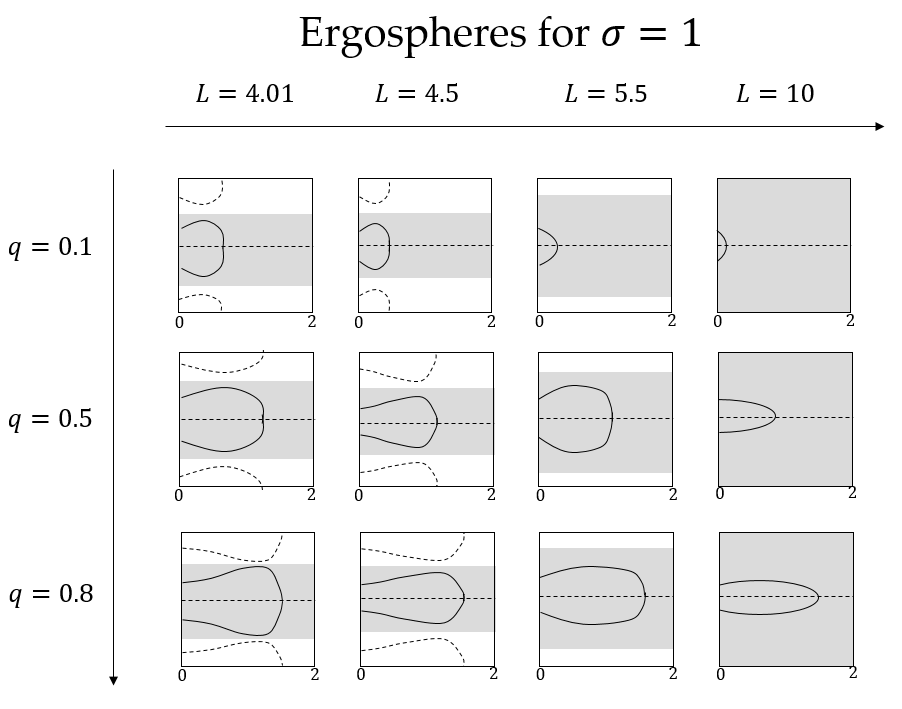}
\caption{Shape  of ergo-sphere depending on $(L,q)$ for fixed $\sigma = 1$. $L$ increases from left to right, $q = \sqrt{1 - p^2}$ increases from top to bottom. The central strip is shown in light-grey; in dashed lines we show the adjacent ergo-spheres.}
\label{Ergospheres_sigma}
\end{figure}

In \autoref{Ergospheres_sigma} we show a typical deformation  of ergospheres for fixed $\sigma$ when the parameter $q$ changes. 

It is interesting to note that there are two mechanisms influencing the shape of the ergosphere that have different effects. 

On the one hand, the higher the rotation of the black hole, the more oblate is the ergosphere. This is the usual effect observed in asymptotically flat Kerr black holes. This deformation is along the $\rho$-direction. 

On the other hand, the smaller is the period $L$, the closer are  horizons to each other, and therefore the bigger the distortion of the horizon. This results in the cigar-like limit of periodic Schwarzschild horizons when the (coordinate) distance between the horizons tends to zero, see \cite{Frolov2003} for further details. This deformation, although not as strong as the previous one (since we are limited by the inequality $4\sigma < L$), deforms the ergosphere along the $z$-direction.

As we can see in \autoref{Ergospheres_sigma}, both deformation take place as $q$ growths and $L$ decreases.

\begin{figure}[h!]
\centering
\includegraphics[scale=0.4]{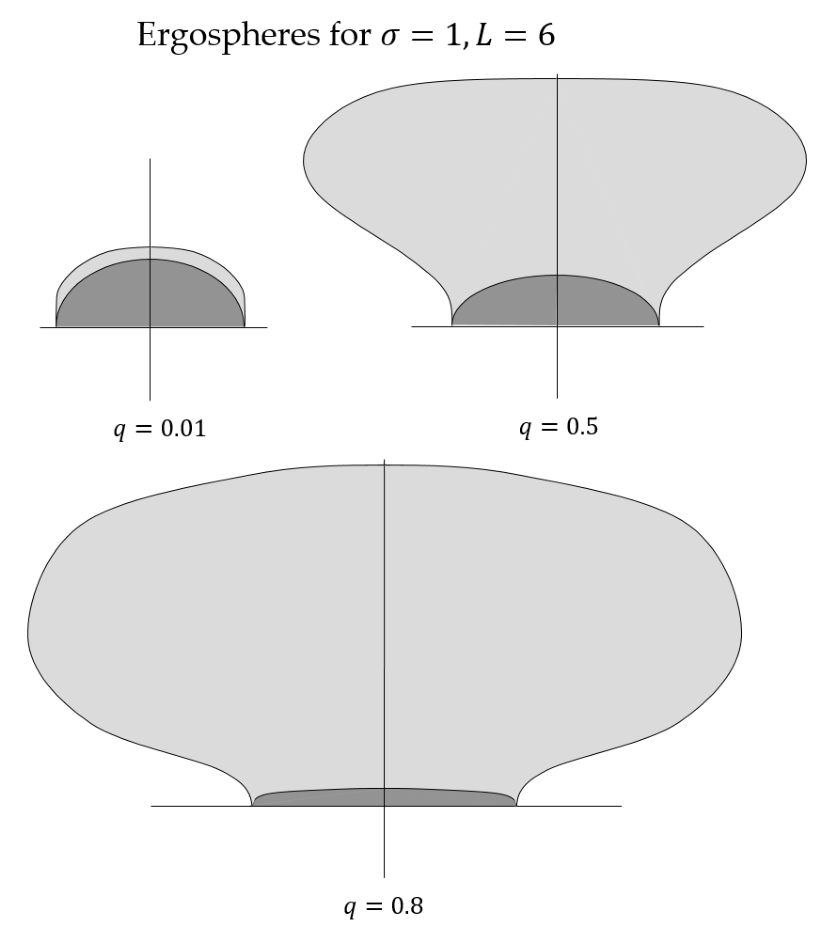}
\caption{Ergosphere isometrically embedded in an Euclidean three-dimensional space for $\sigma= 1$,  $L=6$, and  $q = 0.01, 0.5$ and $0.8$. The horizon embedding is shown in dark grey, while the ergosphere embedding is shown in light-grey.}
\label{Ergospheres_embedding}
\end{figure}

Finally, in \autoref{Ergospheres_embedding} we plot three examples of an ergosphere isometrically embedded in an Euclidean three-dimensional space, fixing $\sigma= 1$ and $L=6$, and taking $q = 0.01, 0.5$ and $0.8$. As a reference, we plot the (also isometrically embedded) shape of the horizon (see \cite{Frolov2003} for a discussion of the isometrically embeddings in the static case).

\section{Outlook} \label{sec_outlook}

In this paper, we explore the parameter space $(\sigma, p ,L)$ corresponding to periodic Kerr black hole solutions, via the Inverse Scattering Method (ISM). The iterative approach of adding solitons, combined with its simplified numerical implementation, allows for a systematic study of the possible configurations in a computationally efficient way. 

As it was already discussed in \cite{Peraza:2022xic} and \cite{Peraza:2024uto}, existence reduces to a concavity test of the $g_{tt}$ component of the metric, which in the Ernst formalism corresponds to the real part the Ernst potential, $\re \E$. In a previous treatment of the problem, \cite{Peraza:2022xic}, numerical solutions were constructed using the harmonic map heat flow, involving a non-trivial analysis and fine tuned boundary conditions for the code to run. Our work simplify considerably the study of the existence since it does not rely on boundary condition for the computation of the concavity of $\re \E$. 

We obtained a detailed diagram of the possible values for $(\sigma, p ,L)$ such that solutions exist, and study the properties of the near horizon geometry, in particular the isometric embeddings into three-dimensional Euclidean space of the horizons and the ergospheres.

While this work has provided numerical evidence for the domain of existence, several questions for future research remain open.

First, the analytical classification of periodic Kerr solutions is still incomplete. A rigorous proof of existence for solutions in the numerically supported parameter regions, particularly in the grey region of Figure \ref{existence_fig}, is a natural extension of this work. Such a proof would require overcoming the challenges posed by the non-flat asymptotic behaviour present in the periodic configurations.

Second, the physical interpretation of periodic Kerr solutions is still missing. Understanding their stability under perturbations, their potential astrophysical implications, and their relationship with higher-dimensional black hole analogs could provide new insights into gravitational physics in both theoretical and practical contexts. 

The study of bounded  null geodesics around the horizon and comparison with known closed photon orbits and photon spheres 
is desirable to understand the geometrical difference between periodic and asymptotically flat cases.

Finally, extending the methods developed here to other families of spacetimes with two or more hypersurface-orthogonal Killing fields — such as those involving cosmological constants or higher-dimensional analogs — could enrich the understanding of gravitational systems in non-asymptotically flat regimes.

\section*{Acknowledgments}

We thank Martin Reiris for useful discussions. JP was partially funded by Fondo Clemente Estable Project FCE\_1\_2023\_1\_175902 and by CSIC Group 883174. Currently funded by a Postdoctoral Fellowship at Concordia University, Montreal. We thank Marco Bertola for numerous discussions and help with numerical analysis of multi-soliton solutions. The work of D.K. was supported in part by the NSERC grant
RGPIN-2020-06816.

\appendix

\section{Kerr solution} \label{Kerr_explicit}

Here we remind how to get
 Kerr solution by adding two solitons to the background solution $\E_0  =1$. Let $\pm \sigma$ be the position of the poles, and let $\alpha_1 = \alpha , \alpha_2 = -\bar{\alpha}$ be  unitary complex constant.

The Ernst potential of the corresponding Kerr solution then looks as follows
$$
\E =   \frac{w(\lambda_1) \alpha_1 - w(\lambda_2) \alpha_2 - \lambda_1 + \lambda_2}{\lambda_1 - \lambda_2 + w(\lambda_1) \alpha_1 - w(\lambda_2) \alpha_2} \;.
$$

Introducing the prolate coordinates, 
$$ 
x = \frac{w(\lambda_1) + w(\lambda_2)}{\lambda_1 - \lambda_2}, \quad y = \frac{w(\lambda_1) - w(\lambda_2)}{\lambda_1 - \lambda_2},
$$
we get 
$$
\E =   \frac{\alpha_1 (x+y) - \alpha_2 (x-y) - 2}{\alpha_1 (x+y) - \alpha_2 (x-y) + 2} \;,
$$
ok
denoting $\alpha = p - i q$ (such that $p^2+q^2=1$), we have \cite{stephani_kramer_maccallum_hoenselaers_herlt_2003}
$$
\E = \frac{px -iqy - 1}{px - iqy +1}\;.
$$

The expression for $\Psi$  looks as follows
\be 
\Psi(\lambda,\xi,\bar{\xi}) = \left( \begin{array}{cc}
\frac{1}{\lambda}  q_0 + q_1  & \frac{1}{\lambda} w(\lambda , \xi) p_0 \\
\frac{1}{\lambda}  w(\lambda , \xi) \bar{p}_0  & \frac{1}{\lambda}  \bar{q}_0 + \bar{q}_1  
\end{array} \right).
\ee
where
\beq 
q_1 &=& 2\frac{px -iqy  }{px -iqy + 1}, \\
p_0 &=& \frac{ - 2}{px -iqy + 1}, \\
q_0 &=& -2\sigma \frac{-iqx + p y }{px -iqy + 1}. 
\eeq

Using the formula \eqref{component_F_metric}, we get
$$
F= F_0 +  \frac{1}{4 (p^2 x^2 + q^2 y^2 - 1)} \left( -4\sigma qp (y^2 -x^2) + \frac{2 \sigma q}{p} (1 - y^2 )px \right)\;.
$$
Choosing $F_0 = -4\frac{\sigma q}{p}$ to guarantee the regularity at the axis outside of the event horizon, we arrive to the standard expression 
$$
F = \frac{2\sigma q}{p} \frac{(1- y^2) (px + 1)}{p^2 x^2 + q^2 y^2 -1}\;.
$$

Furthermore, integrating \eqref{conformal_factor} we get
 
$$ 
e^{2k} =  \frac{ p^2 x^2 + q^2 y^2 -1}{p^2 (x^2 - y^2)}\;.
$$

\providecommand{\noopsort}[1]{}\providecommand{\singleletter}[1]{#1}%

\end{document}